\documentclass[12pt, preprint]{aastex}

\usepackage{graphicx}
\usepackage{amssymb}
\usepackage{layout}
\usepackage{url}
\usepackage{subfigure}
\usepackage{color}
\slugcomment{Submitted to {\it The Astrophysical Journal}}

\shorttitle{Synchrotron Polarization in Blazars}
\shortauthors{H. Zhang, X. Chen \& M. B\"ottcher}

\begin{document}


\title{Synchrotron Polarization in Blazars}


\author{Haocheng Zhang\altaffilmark{1,2}, Xuhui Chen\altaffilmark{3,4} and Markus B\"ottcher\altaffilmark{5,1}}

\altaffiltext{1}{Astrophysical Institute, Department of Physics and Astronomy, \\
Ohio University, Athens, OH 45701, USA}

\altaffiltext{2}{Theoretical Division, Los Alamos National Laboratory, Los Alamos, NM 87545}

\altaffiltext{3}{Institute of Physics and Astronomy, University of Potsdam, 14476 Potsdam-Golm, Germany}

\altaffiltext{4}{DESY, Platanenallee 6, 15738 Zeuthen, Germany}

\altaffiltext{5}{Centre for Space Research, North-West University, Potchefstroom,
2531, South Africa}



\begin{abstract}
We present a detailed analysis of time- and energy-dependent synchrotron polarization signatures
in a shock-in-jet model for $\gamma$-ray blazars. Our calculations employ a full 3D radiation
transfer code, assuming a helical magnetic field throughout the jet. The code considers synchrotron
emission from an ordered magnetic field, and takes into account all light-travel-time and other relevant geometric effects,
while the relevant synchrotron self-Compton and external Compton effects are taken care of with the 2D MCFP code.
We consider several possible mechanisms through which a relativistic shock propagating through the jet may affect
the jet plasma to produce a synchrotron and high-energy flare. Most plausibly, the shock is expected
to lead to a compression of the magnetic field, increasing the toroidal field component and thereby
changing the direction of the magnetic field in the region affected by the shock. We find that such
a scenario leads to correlated synchrotron + SSC flaring, associated with substantial variability in
the synchrotron polarization percentage and position angle. Most importantly, this scenario naturally
explains large PA rotations by $\gtrsim 180^{\circ}$, as observed in connection with $\gamma$-ray flares in
several blazars, without the need for bent or helical jet trajectories or other non-axisymmetric
jet features.
\end{abstract}

\section{Introduction\label{section1}}

Blazars are an extreme class of Active Galactic Nuclei (AGNs). They are known to emit
non-thermal-dominated radiation throughout the entire electromagnetic spectrum, from
radio frequencies to $\gamma$-rays, and their emission is variable on all time scales,
in some extreme cases down to just a few minutes \citep[e.g.,][]{Aharonian07,Albert07}.
It is generally agreed that the non-thermal radio through optical-UV radiation is
synchrotron radiation of ultrarelativistic electrons in localized emission regions
which are moving relativistically (with bulk Lorentz factors $\Gamma \gtrsim 10$) along
the jet. The origin of the high-energy (X-ray through $\gamma$-ray) emission is still
controversial. Both leptonic models, in which high-energy radiation is produced by
the same relativistic electrons through Compton scattering, and hadronic models,
in which $\gamma$-ray emission results from proton synchrotron radiation and emission
initiated by photo-pion-production, are currently still viable \citep[for a review of
leptonic and hadronic blazar emission models, see, e.g.,][]{Boettcher07,BR12,Krawczynski12}.

The radio through optical emission from blazars is also known to be polarized, with
polarization percentages ranging from a few to tens of percent, in agreement with a
synchrotron origin in a partially ordered magnetic field. Both the polarization
percentage and position angle are often highly variable \citep[e.g.,][]{Darcangelo07}.
The general formalism for calculating synchrotron polarization is well understood
\citep[e.g.,][]{Westfold59}, and several authors have demonstrated that the observed
range of polarization percentages and the dominant position angle in blazar jets are
well explained by synchrotron emission from relativistically moving plasmoids in a
jet that contains a helical magnetic field \citep[e.g.,][]{Lyutikov05,Pushkarev05}.
Recently, also the expected X-ray and $\gamma$-ray polarization signatures in leptonic
and hadronic models of blazars have been evaluated by \cite{ZB13}, demonstrating that
high-energy polarization may serve as a powerful diagnostic between leptonic and
hadronic $\gamma$-ray production.

Recent observations of large ($\gtrsim 180^{\circ}$) polarization-angle swings that occurred
simultaneously with high-energy ($\gamma$-ray) flaring activity (\cite{Marscher08,Marscher10,Abdo10}),
have been interpreted as additional evidence for a helical magnetic field structure.
However, on the theory side, there is currently a disconnect between models focusing
on a description of the synchrotron polarization features, and models for the broadband
(radio through $\gamma$-ray) spectral energy distributions (SEDs) and variability. Models
for the synchrotron polarization percentage and position angle necessarily take into account
the detailed geometry of the magnetic field and the angle-dependent synchrotron emissivity
and polarization \citep[e.g.,][]{Lyutikov05}, but typically apply a simple, time-independent
power-law electron spectrum and ignore possible predictions for
the resulting high-energy emission. On the other hand, most models for the broadband SEDs
and variability employ a chaotic magnetic field, where the synchrotron emissivity is angle-averaged,
and any angle dependence of synchrotron and synchrotron-self-Compton emissions (in the co-moving
frame of the emission region) is ignored.

An attempt to combine polarization variability simulations with a simultaneous evaluation
of the high-energy emission, has recently been published by \cite{Marscher14}. In his
Turbulent, Extreme Multi-Zone (TEMZ) model, the magnetic field along the jet is assumed
to be turbulent (i.e., with no preferred orientation), but as electrons in a small fraction
of the jet are accelerated to ultrarelativistic energies when passing through a standing shock,
a variable, non-zero percentage of polarization is expected stochastically from the addition
of synchrotron radiation from a small number of energized cells with individually homogeneous
magnetic fields. While this model does occasionally produce apparent polarization-position-angle
rotations, it seems difficult to establish a statistical correlation between $\gamma$-ray
flaring activity and position-angle swings in this model. More often, it is argued that an
initially chaotic magnetic field is compressed by a shock. As a consequence, in the direction
of the line of sight (LOS), the magnetic field may appear ordered locally \citep[e.g.,][]{Laing80}.
Alternatively, strong synchrotron polarization may result in a model in which the emission region
moves in a helical trajectory \citep[e.g.,][]{VR99} guided by a very strong large scale magnetic
field, so that the magnetic field inside the emission region is very ordered. In this case the
polarization-percentage and position-angle changes can be associated with the motion
of the emission region.

In this paper, we investigate the synchrotron-polarization and high-energy emission signatures
from a shock-in-jet model, in which the un-shocked jet is pervaded a helical magnetic field.
As the shock moves along the jet, it accelerates particles to ultrarelativistic energies. We
consider separately several potential mechanisms through which flaring activity may arise in
such a scenario, including the amplification of the toroidal magnetic-field component. For
the purpose of our simulations, we will employ the time-dependent 2D radiation transfer
model developed by \cite{Chen11,Chen12}. This model assumes an axisymmetric, cylindrical
geometry for the emission region, and uses a locally isotropic Fokker-Planck equation to evolve
the electron distributions. The latest development of this code includes a helical magnetic
field structure to replace the original chaotic structure (with angle-averaged emissivities),
which makes the evaluation of synchrotron polarization possible. However, this evaluation of polarization
requires treatment of synchrotron emission in full 3D geometry. Since there is a large number of free parameters
in this model, we will here focus on a general parameter study, simulating and comparing the
polarization patterns for different possible flaring scenarios, rather than fit the observed
data directly. In a future paper, we plan to combine MHD simulations with this code in order
to constrain the free parameters pertaining to changes in the magnetic-field configuration,
and fit the data directly. We will describe the code setup in \S\ref{section2}, compare different
scenarios in \S\ref{section3}, and discuss the results in \S\ref{section4}.

\section{Code Setup\label{section2}}

In this section, we will first give a brief review of the 2D radiation transfer model
by \cite{Chen11,Chen12}, then introduce the 3D polarization code setup and compare its
result with that of the 2D code.

\subsection{2D Monte-Carlo/Fokker-Planck (MCFP) Code}

\begin{figure}[ht]
\centering
\includegraphics[width=6cm,height=6cm]{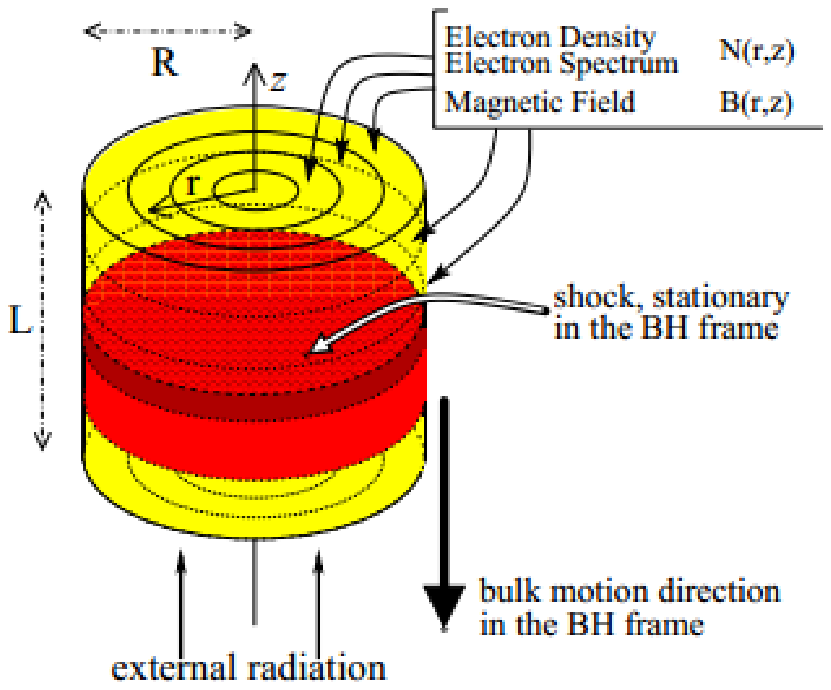}
\hspace{0.5cm}
\includegraphics[width=6.5cm]{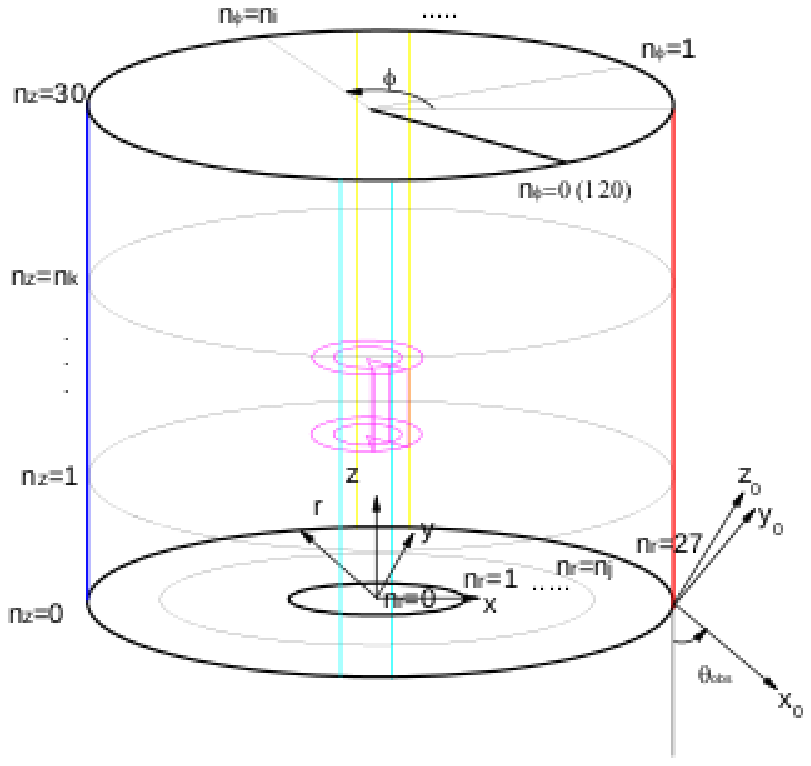}
\vspace{5pt}
\caption{{\it Left:} Sketch of the geometry used in the MCFP code, adapted from \cite{Chen12}.
{\it Right:} Sketch of the geometry (in the co-moving frame) of 3DPol with an enlarged
illustration of one zone (in red) in the emission region. The model uses cylindrical
coordinates, ($r$, $\phi$, $z$), with $n_r$, $n_{\phi}$, $n_z$ being the number of zones in the
respective directions. We define a corresponding Cartesian coordinate system ($x, y, z$) where
$z$ is along the jet axis and the $x$-axis is along the projection of the LOS onto the
plane perpendicular to $z$. The Cartesian coordinates ($x_0, y_0, z_0$) are defined so that
$x_0$ is along the LOS and $z_0$ is the projection of the jet axis onto the plane of the
sky. Both Cartesian coordinate systems are in the co-moving frame of the emission
region. $\theta_{\rm obs}$ is the observing angle between $x_0$ and $z$. Consequently, if
$\theta_{\rm obs} = 90^{\circ}$, ($x, y, z$) = ($x_0, y_0, z_0$). The region between
the two {\it cyan} lines is the zone near the boundary of the emission region in the $-y$
direction, while the {\it yellow} region is offset from the center of the jet in the $+y$
direction. The {\it red} and {\it blue} lines represent the $x=+r_{max}$ (pre-peak)
and $x=-r_{max}$ (post-peak) boundaries of the cylindrical emission region, respectively.
{\it Magenta} region represents a near-central zone in the emission blob.
\label{setup}}
\end{figure}

The code of \cite{Chen11,Chen12} assumes an axisymmetric cylindrical geometry for the emission jet,
which is further divided evenly into zones in radial and longitudinal directions (Fig. \ref{setup}).
The plasma moves relativistically along the jet, which is pervaded by a helical magnetic field,
and encounters a flat stationary shock. In the comoving frame of the emission region,
the shock will temporarily change the plasma conditions as it passes through the jet plasma,
and hence generate a flare. In this paper we consider four parameter changes that may
characterize the effect of the shock: 1. amplification of the toroidal component of the
magnetic field; 2. increase of the total magnetic field strength; 3. shortening of the
acceleration time scale of the non-thermal electrons; and 4. injection of additional non-thermal
electrons. The model uses a locally isotropic Fokker-Planck equation to evolve the electron
distributions in each zone and applies the Monte-Carlo method to track the photons. Hence,
the combined scheme is referred to as a Monte-Carlo/Fokker-Planck (MCFP) scheme. Within
the Monte-Carlo scheme, all light travel time effects (LTTE) are considered.

\subsection{3D Multi-zone Synchrotron Polarization (3DPol) Code}

We employ a similar geometry setup in the 3D multi-zone synchrotron polarization code. As a full 3D
description is necessary to evaluate the polarization, we further divide the emission region evenly
in the $\phi$ direction (Fig. \ref{setup}). The assumption of axisymmetry of all parameters is still
kept. As in the MCFP code, all calculations are performed in the co-moving frame of the emission
region. The LOS is properly transformed via relativistic aberration. In each zone, we project the
magnetic field onto the plane of sky in the comoving frame, and use the time-dependent electron
distribution generated by the MCFP simulation, instead of a simple power-law, to evaluate the polarization
via \citep{RL85}
\begin{equation}
\Pi (\omega) = {P_{\perp} (\omega) - P_{\parallel} (\omega) \over
P_{\perp} (\omega) + P_{\parallel} (\omega) }
\label{Pidef}
\end{equation}
where $\omega$ is the frequency, $P_{\parallel} (\omega)$ and $P_{\perp} (\omega)$ are the radiative
powers with electric-field vectors parallel and perpendicular to the projection of the magnetic field
onto the plane of the sky, respectively. $P_{\parallel} (\omega)$ and $P_{\perp} (\omega)$ are
obtained via integration of the single-particle powers $P_{\parallel}(\omega, \gamma)$
and $P_{\perp} (\omega, \gamma)$ over the electron spectrum $N_e(\gamma)$, e.g.,
$P_{\parallel} (\omega) = \int_1^{\infty} P_{\parallel} (\omega, \gamma) \, N_e(\gamma) d\gamma$.
Since the net electric-field vector is perpendicular to the projected magnetic field on the plane of sky
in the comoving frame, the electric vector position angle, also known as polarization angle (PA),
is obtained for each zone, hence we can obtain the Stokes parameters (without normalization) at
every time step via
\begin{equation}
(I,Q,U) = L_{\nu}*(1,\Pi\cos2\theta_E,\Pi\sin2\theta_E)
\end{equation}
where $L_{\nu}$ is the spectral luminosity at frequency $\nu$, $\Pi$ is the polarization percentage
and $\theta_E$ is the electric vector position angle for that zone.
The code then calculates the relative time delay to the observer for each zone,
so as to take full account of the external LTTE, i.e., the time delay in the observed emission due
to the spatial difference of each zone in the emission region. The internal LTTE, which is introduced
through Compton scattering, is irrelevant for the present discussion of synchrotron polarization.
Since the emission from different zones is incoherent, the total Stokes parameters are then calculated \
by direct addition of the Stokes parameters for each zone from which emission arrives at the observer
at the same time. In a post-processing routine to analyze the 3DPol output, we normalize the total
Stokes parameters at every time step to evaluate the polarization, and transform the result back to
the observer's frame.

\begin{figure}[ht]
\centering
\includegraphics[width=14cm]{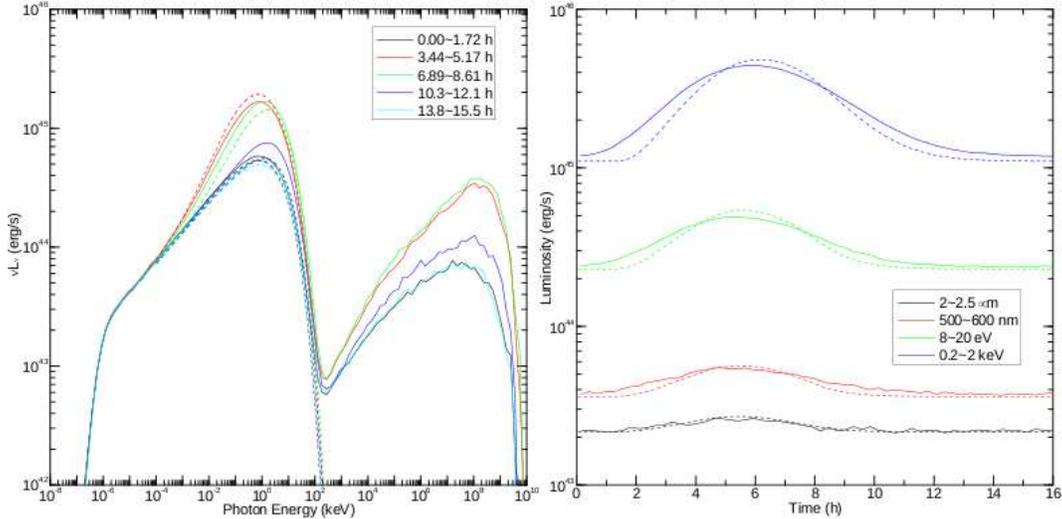}
\vspace{5pt}
\caption{Comparison between the MCFP (solid lines) and the 3DPol (dashed lines) codes.
{\it Left:} The SEDs of Mkn~421 for scenario 4. The time bins are chosen the same for both codes.
{\it Right:} The light curves of Mkn~421 for scenario 4. The energy bins are identical as well.
\label{compare}}
\end{figure}

In order to test our code, we compare the total, time-dependent synchrotron spectra obtained with our
3DPol code, with the result of the MCFP code. This comparison is illustrated in Fig. \ref{compare}.
The overall agreement is excellent, although minor differences can be noticed, which can be attributed
to the fact that the two codes use different ways to treat the radiation transfer. In particular, the
larger number of zones in the 3D geometry used within 3DPol leads to slightly different LTTE
compared to those in the 2D geometry of the MCFP code.

\section{Results\label{section3}}

In this section, we present case studies for two blazars as examples to apply our polarization code:
The high-frequency-peaked BL Lac object (HBL) Mkn~421, which is successfully fitted with a pure SSC
(Synchrotron-self-Compton) model \citep{Chen11}, and the Flat-Spectrum Radio Quasar (FSRQ) PKS~1510-089,
which requires an additional EC (external Compton) component to model the $\gamma$-ray emission. In
the model fit obtained in \cite{Chen12}, the external photons come from a dusty torus. \cite{Chen11,Chen12}
presented model fits to the SEDs and light curves of these two blazars, using their shock-in-jet model
as described above. In our polarization variability study, we will use similar parameters as in
\cite{Chen11,Chen12} and compare the polarization signatures from all potential flaring scenarios
1 - 4, as described above.
In order to facilitate a direct comparison, we choose the same initial
parameters for all scenarios. The parameters are chosen in a way that they
produce adequate flares for both sources in order to allow for direct comparisons and to
mimic the observational data.
Since both MCFP and 3DPol codes are time dependent,
in the beginning there is a period for the electrons and the photons to reach equilibrium,
before we introduce the parameter disturbance produced by the shock. The light curves shown
in all our plots start after this equilibrium has been reached. As the flaring activity in
the four scenarios exhibits different characteristics in duration and in strength, we define
similar phases in the flare development for the purpose of a direct comparison. These phases
correspond approximately to the early flare, flare peak, and late flare, and post-flare (end)
states. As in \cite{Chen11,Chen12}, the ratio between the emission-region dimensions $z$ and
$r$ is chosen to be $\frac{4}{3}$, to mimic a spherical volume.

Due to the relativistic aberration, even though we are observing blazars nearly along the jet
in the observer's frame (typically, $\theta^{\ast}_{\rm obs} \sim 1/\Gamma$, where $\Gamma$ is
the bulk Lorentz factor of the outflow along the jet), the angle $\theta_{\rm obs}$ between LOS
and the jet axis in the comoving frame it is likely much larger. Specifically, if $\theta^{\ast}_{\rm obs}
= 1/\Gamma$, then $\theta_{\rm obs} = \pi/2$. Hence, for our base parameter studies, we set
$\theta_{obs}=90^{\circ}$. This choice turns out to have a considerable effect on the result,
which will be discussed in \S\ref{section41}.

We define the PA in our simulations as follows. $PA=0^{\circ}$ corresponds to the electric-field
vector being parallel to the projection of the jet on the plane of sky. Increasing PA corresponds
to counter-clockwise rotation with respect to the LOS, to $180^{\circ}$ when it is anti-parallel
to the projected jet (which is equivalent to $0^{\circ}$ due to the $180^{\circ}$ ambiguity). In
all runs the zone numbers in three directions are set to $n_z=30$, $n_r=27$ and $n_{\phi}=120$,
which we find to provide appropriate resolution. As is mentioned in \S\ref{section1}, we will only
focus on a parameter study and compare the general flux and polarization features of each scenario.
All results are shown in the observer's frame. Table 1 lists some key parameters.

\begin{table}[ht]
\scriptsize
\parbox{1.0\linewidth}{
\centering
\begin{tabular}{|l|c|c|}\hline
Parameters                                          & \multicolumn{2}{c |}{Initial Condition}    \\ \hline
Source                                              & Mkn~421           & PKS~1510-089           \\ \hline
Bulk Lorentz factor $\Gamma$                        & $33.0$            & $15.0$                 \\ \hline
$Z$ $(10^{16}cm)$                                   & $1.0$             & $8.0$                  \\ \hline
$R$ $(10^{16}cm)$                                   & $0.75$            & $6.0$                  \\ \hline
Size of the shock region in $z$ $(10^{16}cm)$       & $0.1$             & $0.8$                  \\ \hline
Magnetic field $B$ $(G)$                            & $0.13$            & $0.2$                  \\ \hline
Magnetic field orientation $\theta_{B}$             & $45^{\circ}$      & $45^{\circ}$           \\ \hline
Electron density $n_e$ $(10^{2}cm^{-3})$            & $0.8$             & $7.37$                 \\ \hline
Electron minimum energy $\gamma_{min}$              & $10^2$            & $50$                   \\ \hline
Electron maximum energy $\gamma_{max}$              & $10^5$            & $2*10^4$               \\ \hline
Electron spectral index $p$                         & $2.3$             & $3.2$                  \\ \hline
Electron acceleration time-scale $t_{acc}$ $(Z/c)$  & $1.0$             & $0.09$                 \\ \hline
Electron escape time-scale $t_{esc}$ $(Z/c)$        & $0.3$             & $0.015$                \\ \hline
Orientation of LOS $\theta_{obs}$                   & $90^{\circ}$      & $90^{\circ}$           \\ \hline
\multicolumn{3}{c}{}\\
\end{tabular}}

\parbox{.5\linewidth}{
\centering
\begin{tabular}{|l|c|c|}\hline
Parameters                                          & Scenario 1        & Scenario 2             \\ \hline
$B^s/B$                                             & $\sqrt{50}$       & $\sqrt{50}$            \\ \hline
$\theta_{obs}^s$                                    & $84.261^{\circ}$  & $45^{\circ}$           \\ \hline
\multicolumn{3}{c}{}\\ \hline
Parameters                                          & \multicolumn{2}{c |}{Scenario 3}           \\ \hline
Source                                              & Mkn~421           & PKS~1510-089           \\ \hline
$t_{acc}^s/t_{acc}$                                 & $1/5$             & $1/3$                  \\ \hline
\end{tabular}}
\parbox{.5\linewidth}{
\centering
\begin{tabular}{|l|c|c|}\hline
Parameters                                          & \multicolumn{2}{c |}{Scenario 4}           \\ \hline
Source                                              & Mkn~421           & PKS~1510-089           \\ \hline
Inj. $\gamma_{min}$                                 & $10^2$            & $3*10^2$               \\ \hline
Inj. $\gamma_{max}$                                 & $3*10^4$          & $2*10^5$               \\ \hline
Inj. $p$                                            & $1.0$             & $3.2$                  \\ \hline
Inj. rate $(erg/s)$                                 & $5.0*10^{40}$     & $5.0*10^{44}$          \\ \hline
\multicolumn{3}{c}{}\\
\end{tabular}}
\caption{Summary of model parameters. {\it Top:} Initial parameters. Notice that $\gamma_{min}$, $\gamma_{max}$
and $p$ can change before the electrons reach pre-flare equilibrium. {\it Bottom:} Shock parameters for each scenario.
$s$-superscript indicates the parameters during the shock. Scenario 1\&2 choose the same shock parameters for both sources.
Scenario 4 lists the parameters for the injected electrons.}
\end{table}

\begin{figure}[ht]
\centering
\includegraphics[width=15cm]{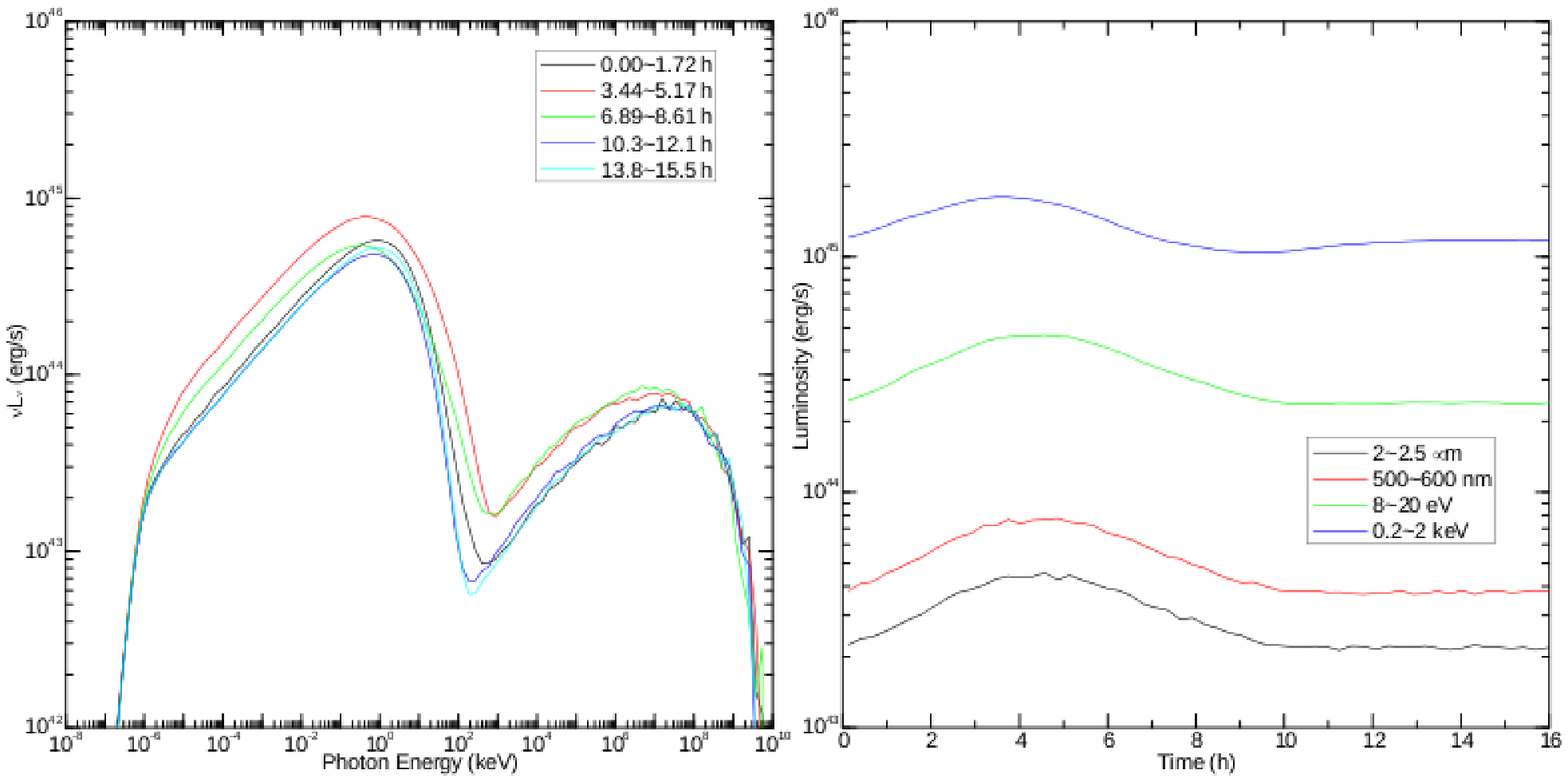}
\vspace{10pt}
\includegraphics[width=15cm]{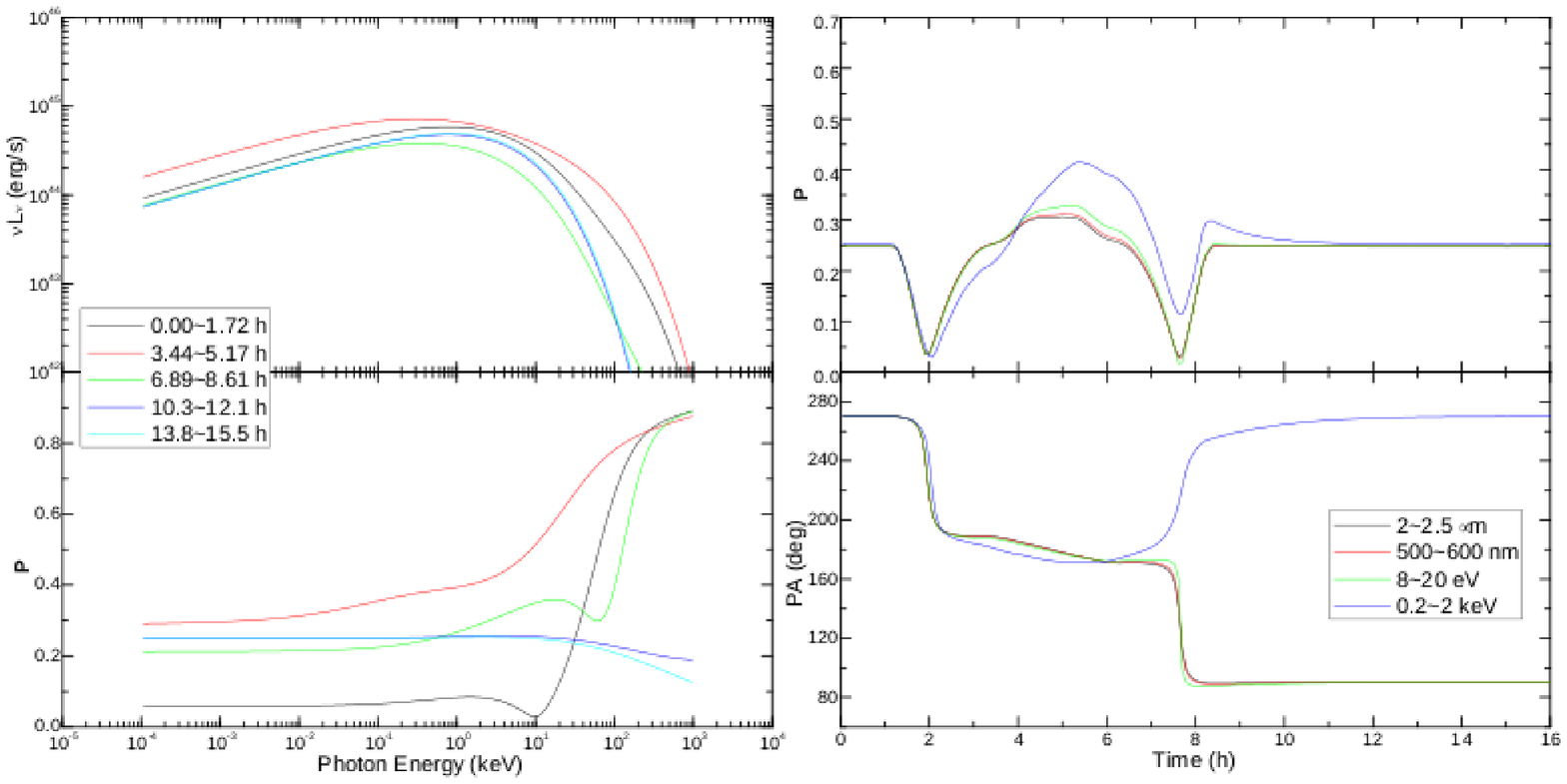}
\vspace{5pt}
\caption{Flaring scenario 1 (change of direction and strength of the magnetic field) for Mkn~421.
{\it Upper left:} The SEDs of Mkn~421 from the MCFP code. SEDs are chosen at approximately the beginning
of the flare (black), peak (red), after peak (green), ending (blue) and back to equilibrium (cyan),
with the same time bin size.
{\it Upper right:} The light curves of Mkn~421 from the MCFP code, chosen at infrared (black),
optical V band (red), UV (green) and soft X-ray (blue) frequencies.
{\it Lower left:} The synchrotron SEDs ({\it top}) of Mkn~421 from the 3DPol code, and the polarization
percentage vs. photon energy ({\it bottom}). The time bins are chosen the same as SEDs from MCFP.
{\it Lower right:} The polarization percentage vs. time ({\it top}), and the polarization position
angle vs. time ({\it bottom}). The energy bands are chosen the same as the light curves from MCFP.
\label{Mkn4215}}
\end{figure}

\subsection{Change of the Magnetic Field Orientation\label{section31}}

In this scenario, the shock instantaneously increases the toroidal magnetic-field component at its location,
so as to increase the total magnetic field strength and change its orientation in those zones. The new
magnetic field will be kept until the shock moves out of the zone; at that time, it reverts back to its
original (quiescent) strength and orientation due to dissipation.

For Mkn~421, since both synchrotron and SSC are proportional to the magnetic field strength, we see flares
in both spectral bumps, though the $\gamma$-ray flare has a much lower amplitude (Fig. \ref{Mkn4215}).
It is also obvious that
the polarization percentage has a dependence on the photon energy, although patterns above $\sim\!10$~keV
are resulting from the electron distribution cut-off. In addition, it also has a time dependency. It is
interesting to note that, unlike the light curve, which is symmetric in time, the polarization percentage
has an asymmetric time profile, especially for higher energies. Furthermore, the polarization
angles are shown to have $\sim\!180^{\circ}$ swings, although the X-ray polarization angle reverts back
to its original orientation after the initial $\sim 90^{\circ}$ rotation, instead of continuing to rotate
in the same direction, as in the lower-frequency bands. As we will elaborate in detail below, all these
phenomena can be explained as the combined effect of electron evolution and LTTE.

\begin{figure}[ht]
\centering
\includegraphics[width=15cm]{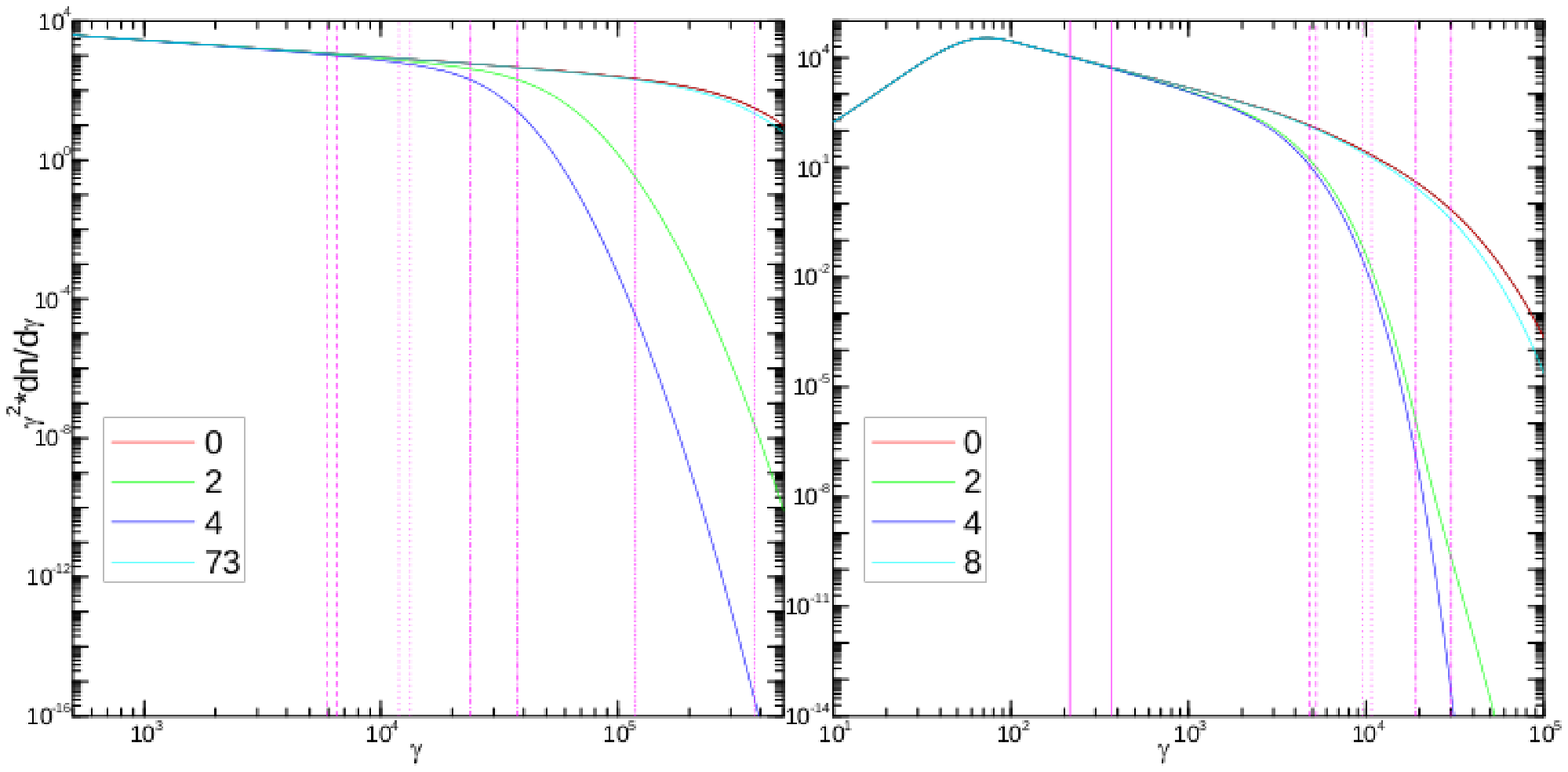}
\vspace{5pt}
\caption{The electron spectra chosen at different time steps (code unit) for scenario 1 for Mkn~421 ({\it left})
and for PKS~1510-089 ({\it right}). $0$: shock turns on (identical to the pre-shock equilibrium,
since the electrons have no time to evolve);
$2$: in the middle of the shock; $4$: shock just turns off; and the final time step which
varies by scenarios and sources is when the electron has evolved to the post-shock equilibrium (although given
enough time, electrons will evolve back to the pre-shock equilibrium, but that process is extremely slow
and not relevant to both luminosity and polarization). The region between magenta vertical lines represent
the electron energies that correspond to the photon energies we choose in the light curves and
polarization vs. time plots. Dashed is infrared, dotted is optical, dashed-dotted is UV, solid is radio
(PKS~1510-089 only), short-dashed is X-rays (Mkn~421 only).
\label{Electron5}}
\end{figure}

Since we assume that every zone in the jet has identical initial conditions and the shock will affect
the same change everywhere, we can simply choose one zone to represent the electron evolution of the
emission region (strictly speaking, due to internal LTTE and other geometric effects,
different zones will be subject to slightly different SSC cooling rates; however, we have carefully
checked the electron spectra and found this effect to be negligible in the cases studied here).
This is illustrated in Fig. \ref{Electron5}. When the shock reaches the zone, it will
increase the magnetic field strength, hence synchrotron cooling becomes faster. Therefore the electron
spectrum becomes softer while the shock is present, especially at higher electron energies, resulting
in a higher possible maximal polarization percentage ($\Pi_{max}=\frac{p+1}{p+7/3}$, where $p$ is the
local electron spectral index in the energy range responsible for the synchrotron emission at a given
frequency, see \cite{RL85}). After the shock leaves the zone, the electrons gradually evolve back to
equilibrium. This process takes longer at higher energies, hence the polarization percentage for more
energetic photons recovers more slowly. Nevertheless, the X-ray light curve appears to evolve
faster than at the lower-frequency ones. The reason for this is that the flare amplitude (compared to
the equilibrium emission) is so low at X-ray frequencies, that even if the electrons have not yet reached
equilibrium near the end of the flare, their contribution to the total synchrotron flux is negligible.

\begin{figure}[ht]
\centering
\includegraphics[width=15cm]{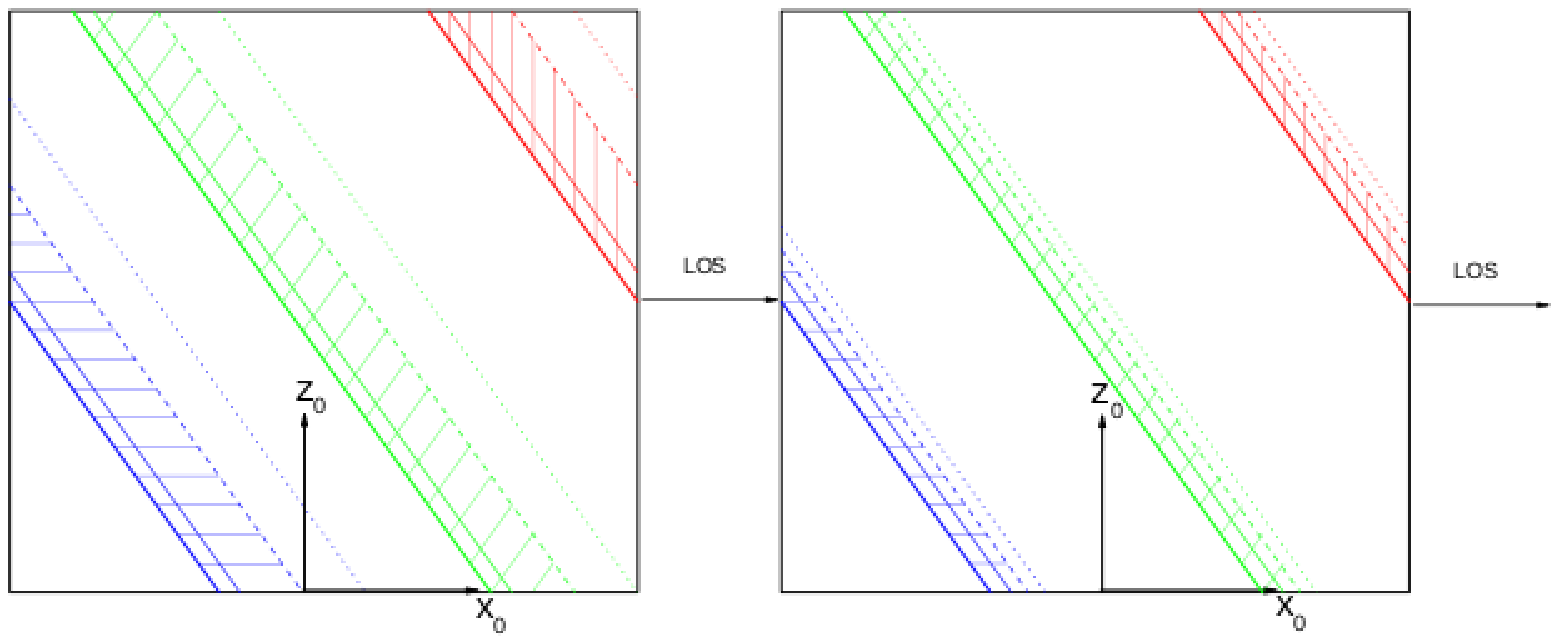}
\vspace{5pt}
\caption{Sketches of a vertical slice of the cylindrical emission region at different times.
{\it Left:} Sketch for Mkn~421. {\it Right:} Sketch for PKS~1510-089. The
LOS is assumed to be directed in $+x$. The shock propagates through the jet from top to
bottom. Solid lines demarcate regions in the jet where the shock is present at equal photon-arrival
times at the observer. {\it Red} (approximately {\it red} in Fig. \ref{setup}) corresponds to
the pre-peak observer time, {\it green} (approximately {\it cyan} and {\it yellow} in Fig. \ref{setup})
represents the flare peak and {\it blue} (approximately {\it blue} in Fig. \ref{setup})
the post-peak time. The flaring region is between the bold and the thin solid lines,
which correspond to the zones where the shock
is currently present. The shaded region between the bold solid line
and the dashed line is what we call the polarization region, containing the flaring region and the evolving
region of recently shock-accelerated electrons. The dotted line represents the zones where electrons have
evolved to the post-flare equilibrium. Although the electrons in the region between the dashed line and the
dotted line are still evolving, their contribution to the polarization is negligible. Hence all points outside
the polarization region are called the background region, or the non-flaring region.
\label{electronsketch}}
\end{figure}

\begin{figure}[ht]
\centering
\includegraphics[width=15cm]{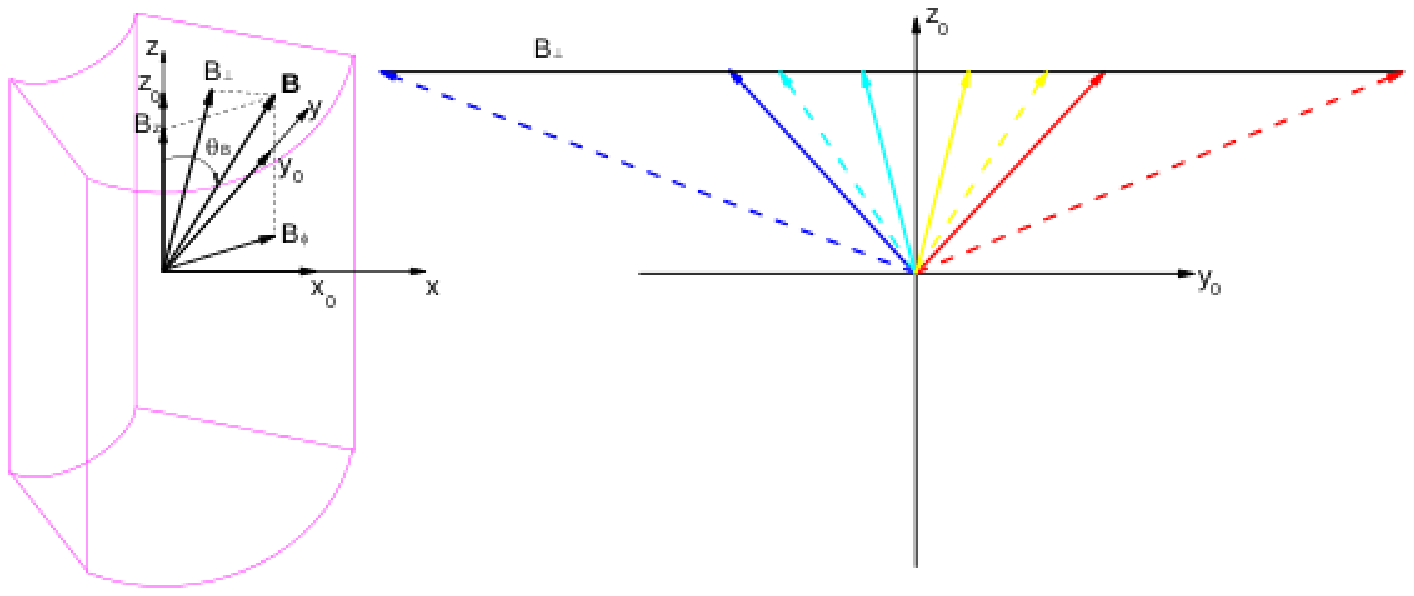}
\vspace{5pt}
\caption{Sketches of magnetic field in the emission blob for $\theta_{obs}=90^{\circ}$.
{\it Left:} Sketch for the magnetic field in the {\it magenta} zone shown in Fig. \ref{setup}.
The coordinates are illustrated in
Fig. \ref{setup}. The total magnetic field ${\bf B}$ in that zone is assumed to have only two components,
$B_z$ and $B_{\phi}$; $\theta_B$ is the angle between ${\bf B}$ and the $z$-axis.
$B_{\perp}$ is the projection of ${\bf B}$ on the plane of sky ($y_0, z_0$).
{\it Right:} Sketch for $B_{\perp}$ at different locations in the emission region.
{\it Red}, {\it yellow}, {\it cyan} and {\it blue} correspond to the regions shown in Fig. \ref{setup}.
When $\theta_{obs}=90^{\circ}$, the $z_0$ component of $B_{\perp}$, $B_{z_0}=B_z$ throughout the emission blob,
while $B_{y_0}\leq B_{\phi}$, and equality is obtained at the $x=+r_{max}$ ({\it red}) and $x=-r_{max}$
({\it blue}) boundaries. The solid arrows represent the initial magnetic field orientation ($\theta_B=45^{\circ}$),
while the dashed arrows illustrate the magnetic field orientation change in scenario 1 ($\theta_B\sim\!84.3^{\circ}$).
\label{bsketch1}}
\end{figure}

We now discuss the influence of LTTEs on the polarization signatures. The situation is
illustrated in Fig. \ref{electronsketch}. In equilibrium, the Stokes parameter $U$
will cancel out because of the axisymmetry. Additionally, despite the fact that
$\theta_B=45^{\circ}$ implies $B_{\phi} = B_z$,
the ``effective toroidal component'', $B_{y_0}$, i.e., the $y_0$ component of $B_{\perp}$
on the plane of sky in the comoving frame, which is generally a fraction of $B_{\phi}$,
is lower than the ``effective poloidal component'', which is equal to $B_z$.
Thus, the polarization is dominated by
$B_z$. Therefore at the initial state, the polarization percentage is relatively low, and the
polarization position angle is at $270^{\circ}$ (or $90^{\circ}$, considering the $180^{\circ}$
ambiguity). However, when the shock reaches the emission region, $B_{\phi}$ begins to dominate.
Due to LTTE, the observer will initially only see the $+x$ side of the flaring region
(Fig. \ref{electronsketch}), which has a preferential magnetic field predominantly in the $+y_0$
direction (Figs. \ref{electronsketch}, \ref{bsketch1}).
Since the emission from the flaring region is much stronger
than other parts of the emission blob, it will quickly cancel and then dominate over the polarization caused
by $B_z$ in the background region. Hence the polarization percentage will first drop to almost zero
and then rapidly increase, while the PA will drop to $\sim\!180^{\circ}$, representing an electric-field vector
directed along the jet, caused by the dominant $B_{y_0}$. Furthermore, the electron spectrum evolves
relatively slowly and the cooled high energy electrons give rise to very high $\Pi_{max}$. Therefore,
the observed polarization will have contributions not only from the flaring region, but also from zones
with more evolved electron distributions, where the shock has passed recently. We can observe in
Fig. \ref{electronsketch} that this ``polarization region'' is not symmetric from pre-peak to post-peak,
resulting in an asymmetry in time, especially at higher energies.

There is, however one additional factor. We can see in the light curves (Fig. \ref{Mkn4215}) that the
X-ray flare-to-equilibrium ratio is much smaller than at lower energies. Hence, in X-rays, the flaring
region takes longer to dominate the polarization patterns, and they revert back to equilibrium faster.
Furthermore, high energy electrons take longer to evolve back to equilibrium, while still providing a
considerable $\Pi_{max}$. For this reason, the X-ray polarization region will be much larger than at
lower energies. Therefore, when the flaring region moves to $-x$ in Fig. \ref{electronsketch},
where the preferential magnetic field is directed in $-y_0$, the X-ray photons will
be dominated by the evolving and the background region on the $+x$ side,
although lower energy photons are still dominated
by the flaring region. This gives rise to the interesting phenomenon that at lower energies the PA
will continuously drop to $90^{\circ}$ (which is equivalent to $270^{\circ}$ because of the
$180^{\circ}$ ambiguity) as the polarization region gradually moves to $-x$ and out of the
emission region; while for X-rays it instead reverts back, as the evolving region on the $+x$ side dominates the
polarization, causing the magnetic field again to be preferentially oriented in the $+y_0$ direction,
mimicking the pre-peak situation.

\begin{figure}[ht]
\centering
\includegraphics[width=15cm]{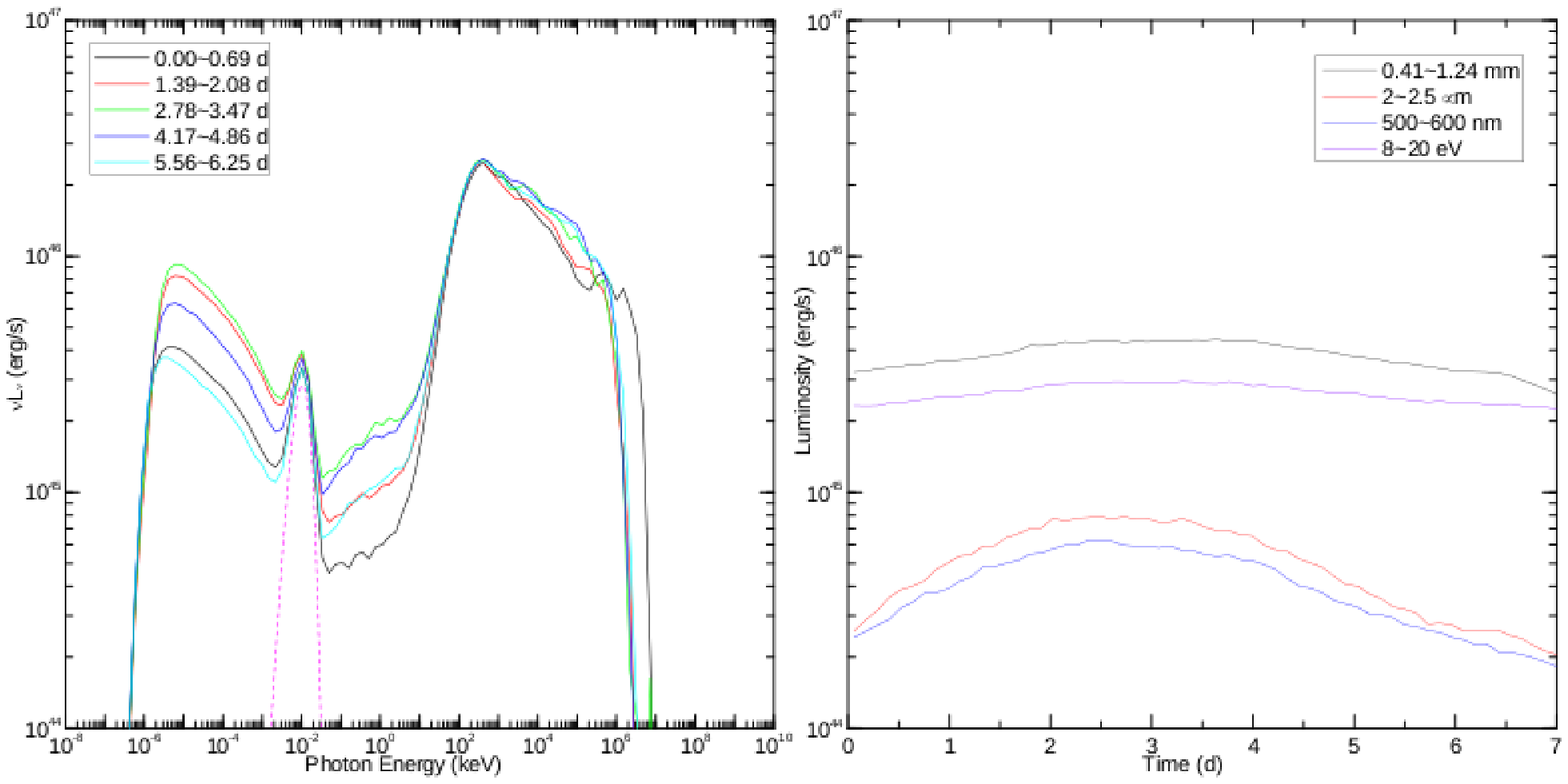}
\vspace{10pt}
\includegraphics[width=15cm]{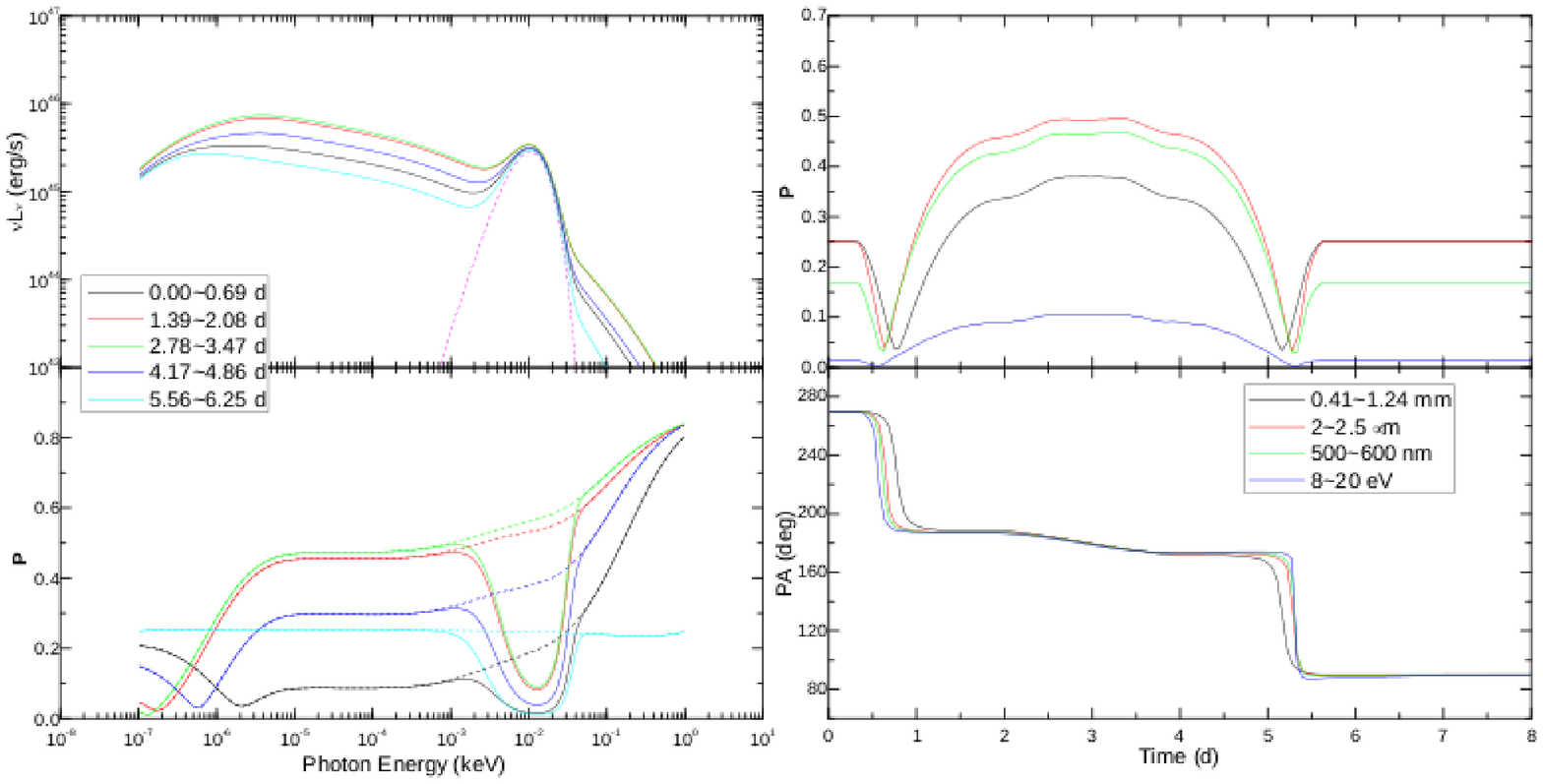}
\vspace{5pt}
\caption{Flaring scenario 1 (change of direction and strength of the magnetic field)
for PKS~1510-089.
{\it Upper left:} SEDs of PKS~1510-089 including the external photon field contribution,
at approximately the beginning of the flare (black), before peak (red), peak (green), after peak (blue)
and back to equilibrium (cyan), with the dashed line for the external photon field contribution.
{\it Upper right:} The light curves including the external photon field contribution,
at radio (black), infrared (red), optical V band (green), UV (blue).
{\it Lower left:} The synchrotron SEDs, including the external photon field, from 3DPol
({\it top}), with the dashed line for the external photon field contribution; and the polarization
percentage vs. photon energy with the external photon contamination considered ({\it bottom}), where dashed
lines represent the polarization percentage without the contamination. {\it Lower right:} The polarization
percentage vs. time with external contamination ({\it top}), and PA vs. time ({\it bottom}).
\label{Pks15105}}
\end{figure}

PKS~1510-089 presents a similar situation, although there are some major differences. First, PKS1510-089
requires a dominating EC component at $\gamma$-ray energies, which is independent of the magnetic field
strength. Thus in the current scenario, no flare is visible in the Compton bump (Fig. \ref{Pks15105}).
Also, due to the contamination of the external thermal radiation from the dust torus, which is unpolarized,
the observed polarization percentage will be considerably diminished. Furthermore, EC also introduces strong
cooling to the electrons, leading to softer electron spectra (Fig. \ref{Electron5}). As a result, $\Pi_{max}$
is much higher. Therefore at the flare peak, the polarization percentage is much higher than that in Mkn~421.
Additionally, the relative electron evolution rate is faster (although the total flare time is longer than
that of Mkn~421, due to the larger dimensions of the emission region). Consequently, the polarization region
is much smaller, nearly equivalent to the flaring region (Fig. \ref{electronsketch}). This will make the
polarization region highly symmetric from pre-peak ($+x$) to post-peak ($-x$), so that the time asymmetry
found in Mkn~421 is not present in the case of PKS~1510-089. Additionally, at the flare peak the polarization
region itself will concentrate on the central region ({\it green}) in Fig. \ref{electronsketch},
where $B_{y_0}$ is much weaker. Thus, the polarization dominated by the effective toroidal component
will be diminished to a certain extent, creating a plateau at the flare peak, which is much lower than
$\Pi_{max}\simeq\!0.75$. In fact, we also find a similar but weaker effect in Mkn~421, due to a larger
polarization region; at X-rays, however, the very large polarization region suppresses this effect, which is
why the X-ray polarization percentage exhibits a pronounced peak.

Comparing the predicted PA swings in Figs. \ref{Mkn4215} and \ref{Pks15105} to the ones observed in several
blazars, in connection with $\gamma$-ray flaring activity, one notices that the observed PA swings are more
gradual than the ones predicted here. However, we remind the reader that we employ the simple assumption that
the magnetic field is instantaneously changed by the shock. In reality, this change will occur over a finite
amount of time. Therefore, with a more realistic time profile of the magnetic-field change, the predicted PA
swing will be much smoother, instead of two rapid drops and a plateau in between. Note also that rather
step-like PA rotations, similar to the features found in our simulations, have in fact been observed in
S5~0716+714 by \cite{Ikejiri11}. We also point out that the $\sim\!180^{\circ}$ PA swing we show here is
the result of one individual disturbance moving through the jet. If there are multiple disturbances (flares)
in succession, the PA will rotate up to $180^{\circ}$ times the number of flares.

There is an ambiguity in the helical magnetic field handedness. In our model setup, we chose
it to be right-handed and against the bulk motion direction. If it were left-handed, the PA rotation would appear
to be in the opposite direction, but everything else would remain the same.
Also notice that even the light curves will not be
symmetric in time because of the asymmetry in time between the dynamics of the shock moving through
the emission region and the electron cooling. However, in cases where the size of the active region,
energized by the passing shock, is much smaller than the overall jet emission region
(e.g., due to dominant EC cooling, the time scale for electron evolution in PKS~1510-089
is much shorter than in the case of Mkn421), this effect is
minor, yielding nearly symmetric light curves.

The $\gamma$-ray emission from PKS~1510-089 is due to
EC, for which changes in the synchrotron component are irrelevant, while the light curve features in our code
are identical to those resulting from the MCFP code, as presented in \cite{Chen11,Chen12}. In the discussion
of the following scenarios, we will show the time-dependent SEDs and light curves only for Mkn~421,
and restrict the discussion of PKS~1510-089 to the polarization signatures.

\begin{figure}[ht]
\centering
\includegraphics[width=15cm]{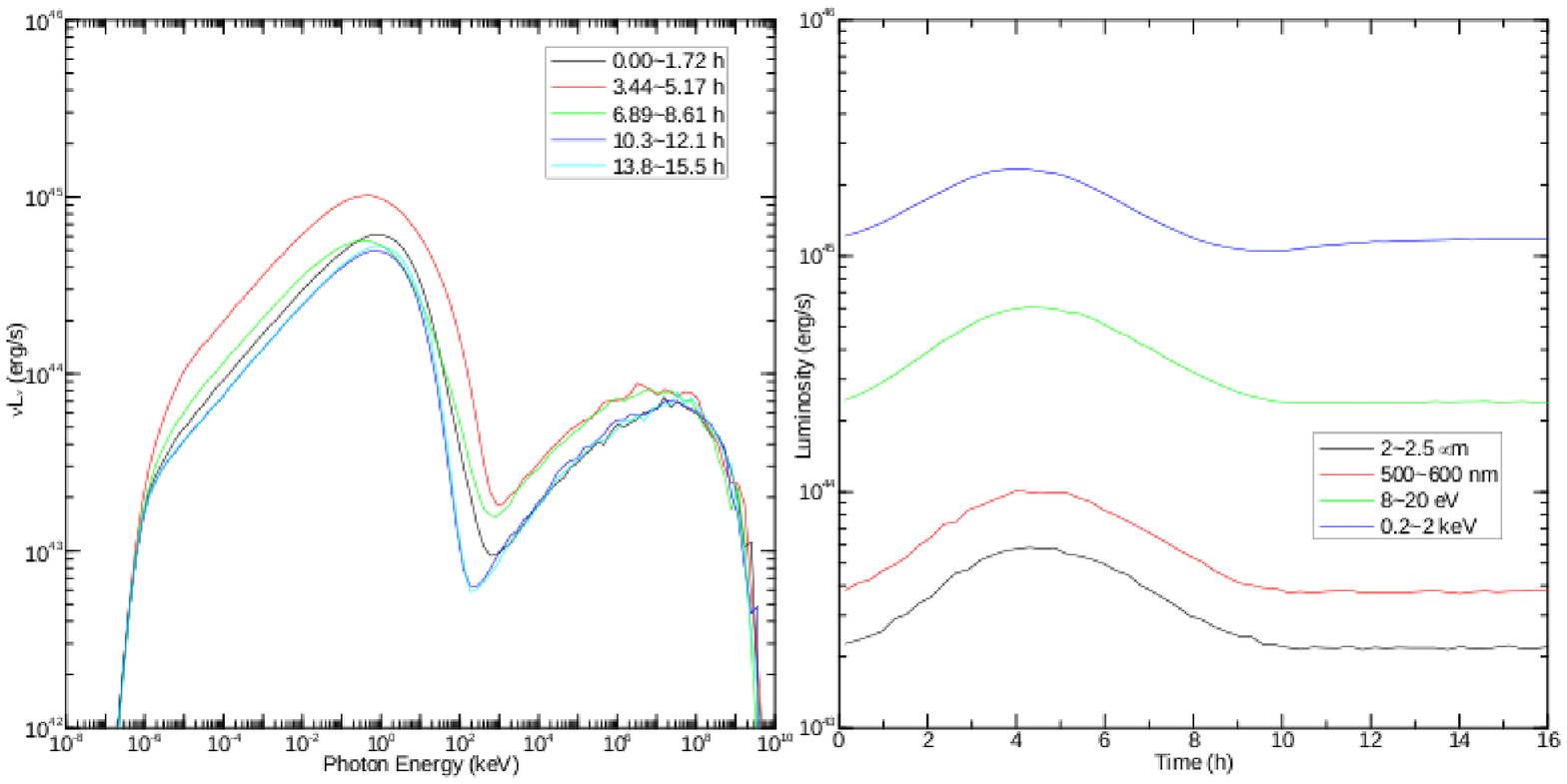}
\vspace{10pt}
\includegraphics[width=15cm]{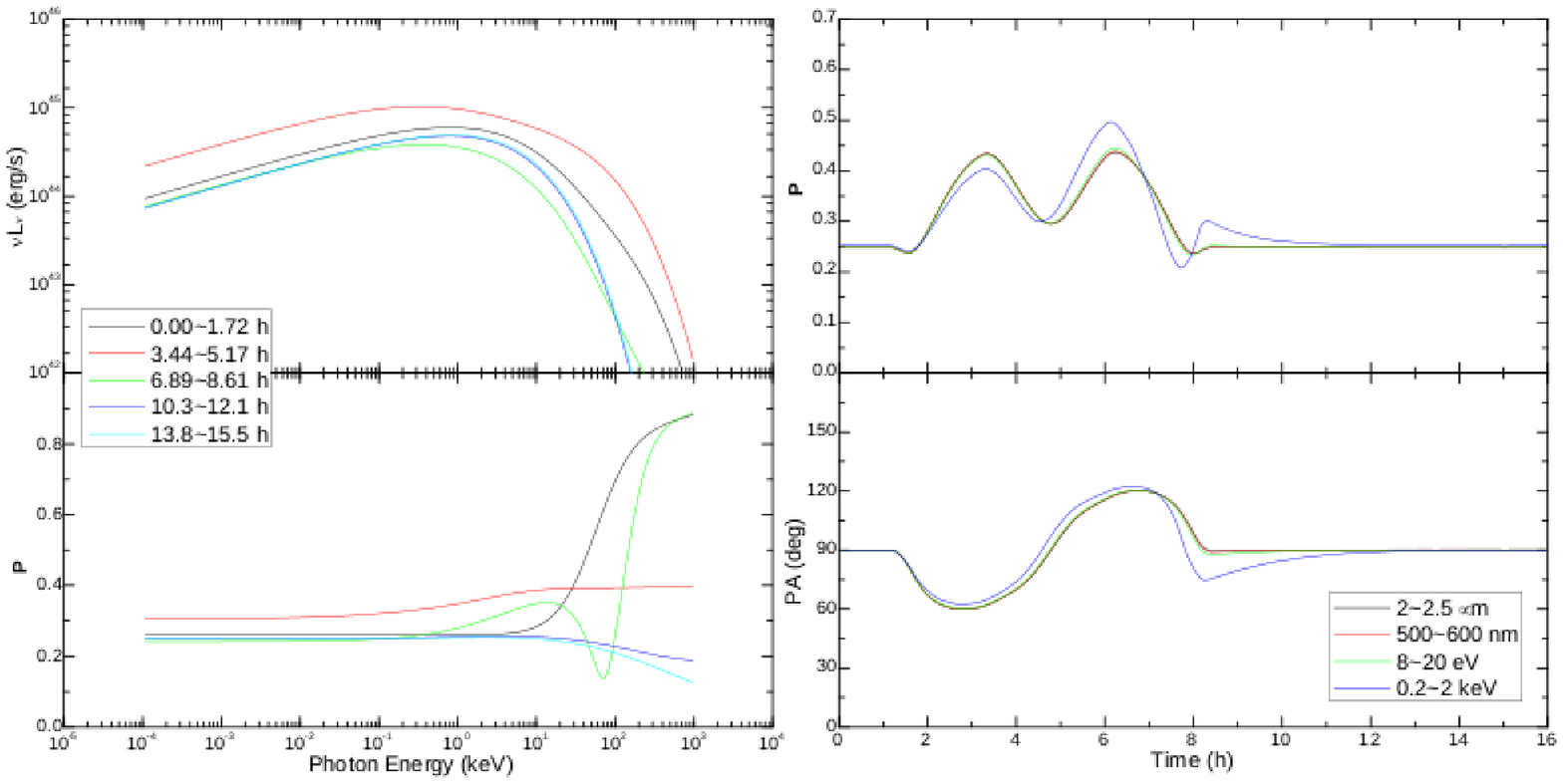}
\vspace{5pt}
\caption{Flaring scenario 2 (increasing magnetic-field strength with unchanged orientation) for
Mkn~421. Panels and line styles are as in Fig. \ref{Mkn4215}.
\label{Mkn4213}}
\end{figure}

\subsection{Increase of the Magnetic Field Strength}

In this scenario, we assume that the shock only increases the total magnetic field strength
at its location, leaving its orientation unchanged. Since the electron evolution is independent
of the magnetic field orientation, it appears identical to the above scenario (Fig. \ref{Electron5}).
The same applies to the SEDs and light curves. However, the polarization patterns in time show
major differences. Since the magnetic field is oriented at $45^{\circ}$ to the $z$-axis, $B_{\phi}$
and $B_z$ will be equal throughout the emission region. Due to axisymmetry, $B_{\perp}$ in the
polarization region will be confined in a cone of $\pm 45^{\circ}$. Therefore, the polarization
induced by $B_z$, is always dominant, so that
the PA will be confined to at most $(45^{\circ},135^{\circ})$. Since also the polarization of the
background region is dominated by the effective poloidal component, these two will add up, resulting in a
slightly higher maximal polarization percentage compared to scenario 1. However, in the immediate
neighborhood of the starting and ending points of the flare, the situation is a little bit
different: although the effective toroidal component $B_{y_0}$ is still weaker than $B_z$,
the two components are closer in magnitude. Hence $B_{\phi}$ will diminish the polarization
percentage by a small amount. Thus the two sharp dips shown in the previous polarization
percentage patterns become much smaller.

\begin{figure}[ht]
\centering
\includegraphics[width=15cm]{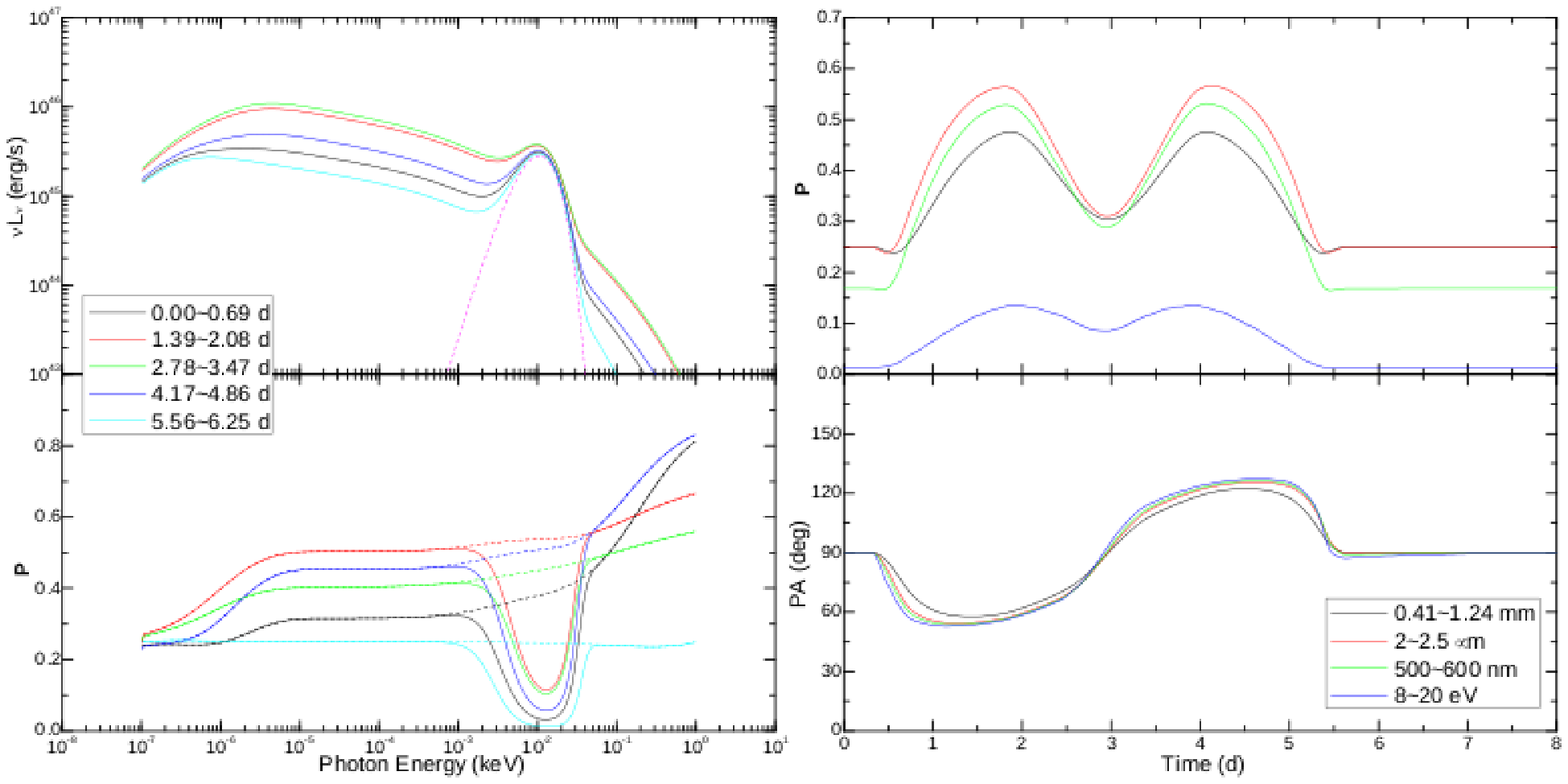}
\vspace{5pt}
\caption{Flaring scenario 2 (increasing magnetic-field strength with unchanged orientation) for
PKS~1510-089. Panels and line styles are as in the bottom two panels of Fig. \ref{Pks15105}.
\label{Pks15103}}
\end{figure}

After that, again due to LTTE, only the $+x$ side of the flaring region is observed initially,
which has a preferential magnetic field oriented at $\sim\!-135^{\circ}$ (Fig. \ref{bsketch1}).
Therefore the polarization is dominated by Stokes parameter $U$. Hence we observe that the
polarization percentage increases and the PA moves to $45^{\circ}$. However, a basin forms
at the flare peak, replacing the previous plateau, and the PA moves back to $90^{\circ}$.
This is because the polarization region at the flare peak (${\it green}$ in
Figs. \ref{electronsketch}, \ref{bsketch1}) is dominated by $B_z$,
while the background region on the $-x$ and $+x$ side is just like the initial state.
Therefore, the Stokes parameter $U$ contributions will cancel out due to axisymmetry,
leaving the polarization dominated by $B_z$. The same applies when the polarization region
moves to the post-peak position ({\it blue} in Fig. \ref{electronsketch}).
Slight differences in the X-ray behavior are again explained by the slower electron
evolution back to equilibrium.

\begin{figure}[ht]
\centering
\includegraphics[width=15cm]{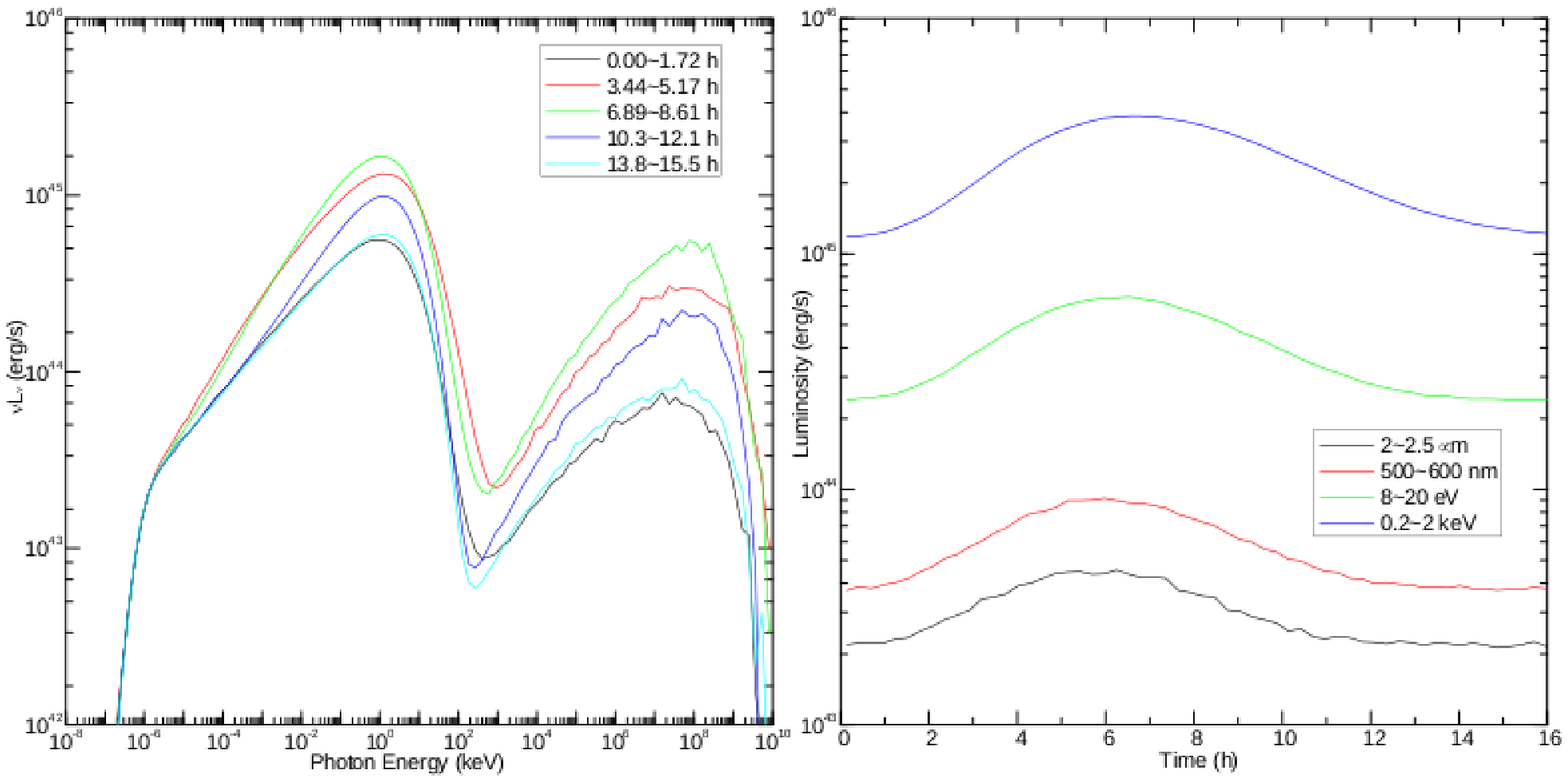}
\vspace{10pt}
\includegraphics[width=15cm]{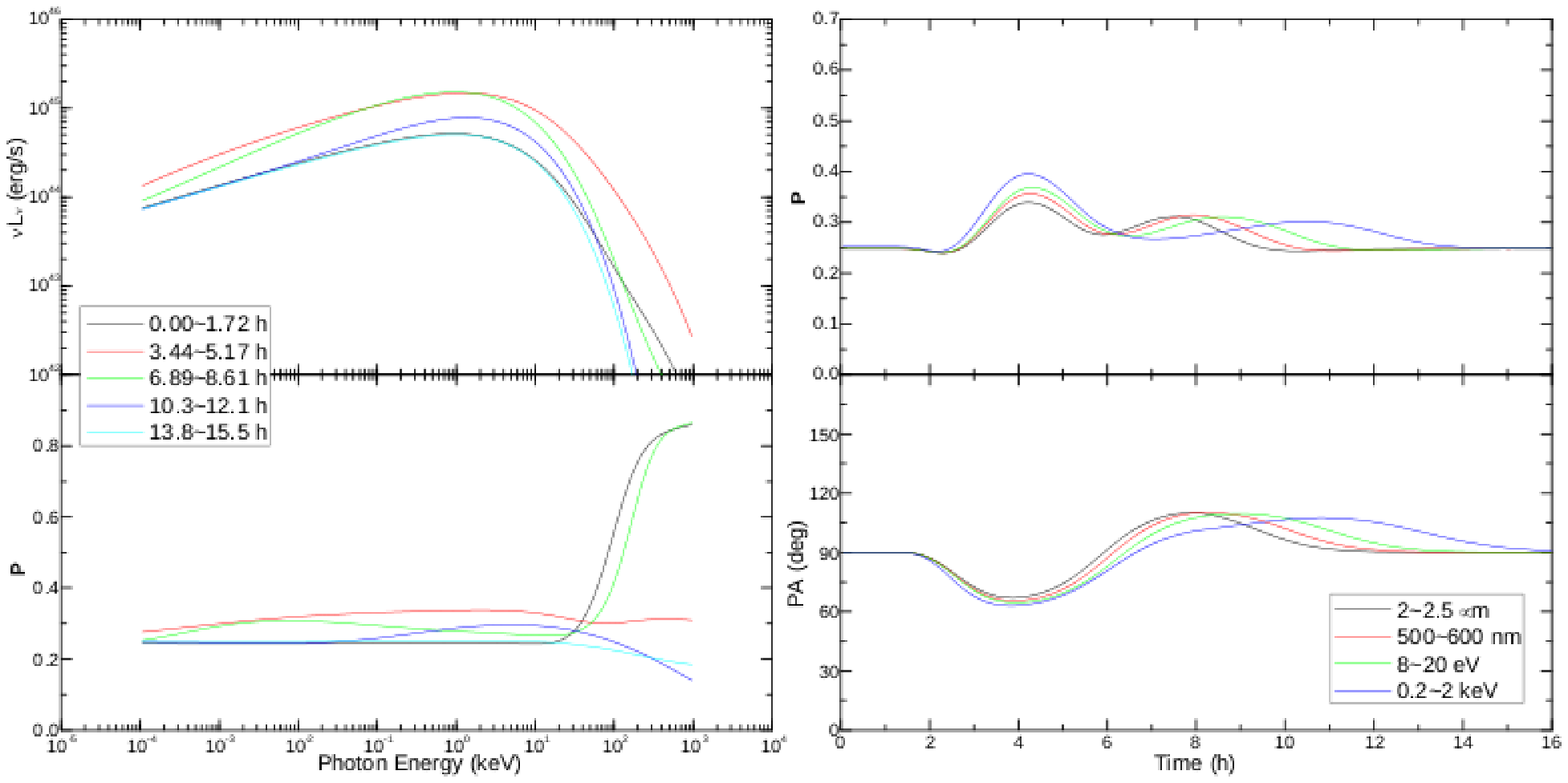}
\vspace{5pt}
\caption{Scenario 3 (shortened acceleration timescale) for MKN~421. Panels and line styles are as
in Fig. \ref{Mkn4215}.
\label{Mkn4214}}
\end{figure}

\subsection{Shortening of the Acceleration Time Scale}

\begin{figure}[ht]
\centering
\includegraphics[width=15cm]{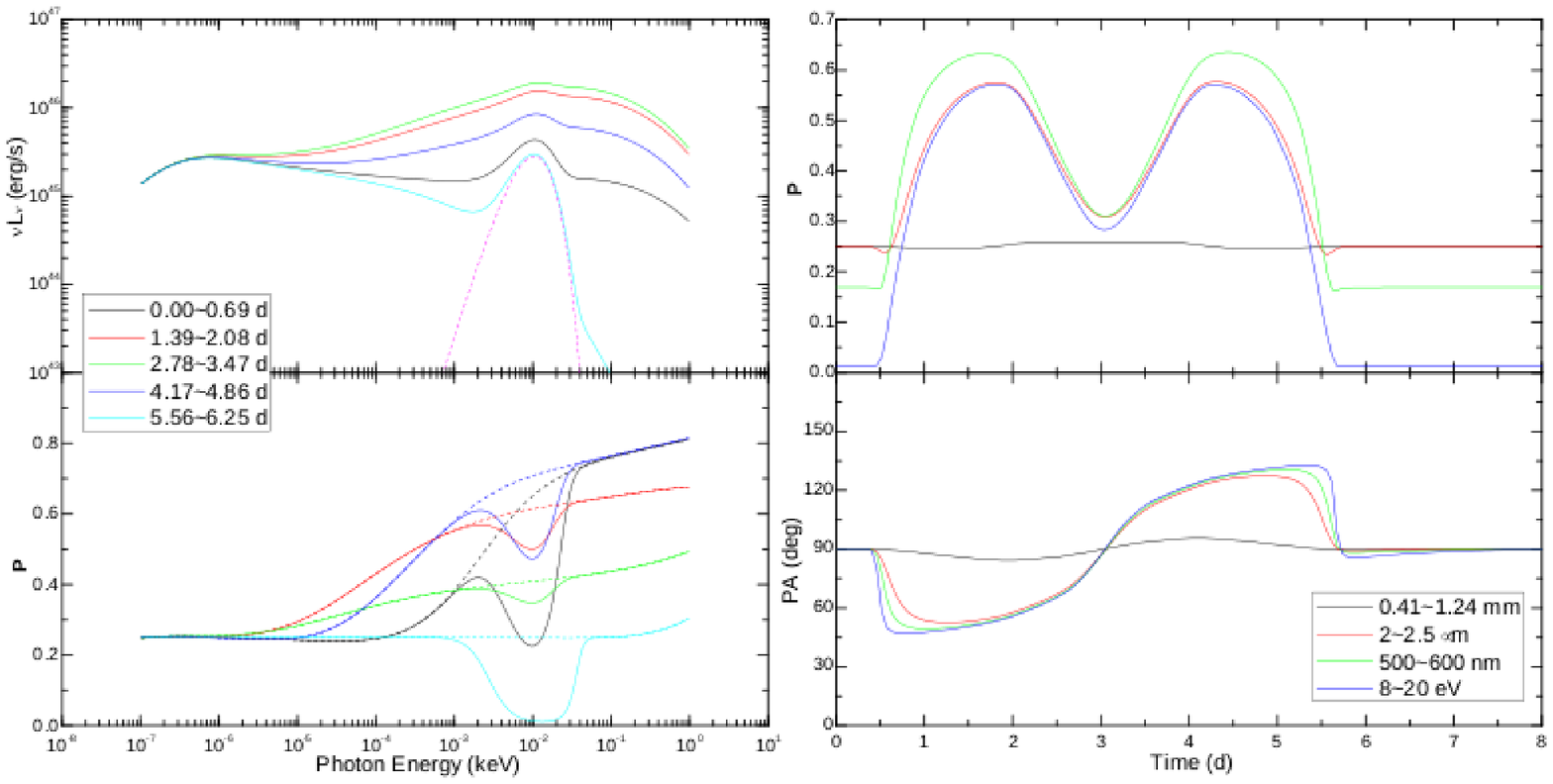}
\vspace{5pt}
\caption{Scenario 3 (shortened acceleration timescale) for PKS~1510-089. Panels and line styles are as
in Fig. \ref{Pks15103}.
\label{Pks15104}}
\end{figure}

In this scenario, the shock is assumed to lead to more efficient particle acceleration by
instantaneously shortening the local acceleration time scale. As a result, the electrons
will be accelerated to higher energy, leading to flares in both synchrotron and Compton
emission (Figs. \ref{Mkn4214}, \ref{Pks15104}). At the same time, the peaks of both
spectral components move to considerably higher energies. However, since the magnetic
field orientation remains unchanged, as in the previous scenario, the PA will stay
confined to at most $(45^{\circ},135^{\circ})$.

\begin{figure}[ht]
\centering
\includegraphics[width=15cm]{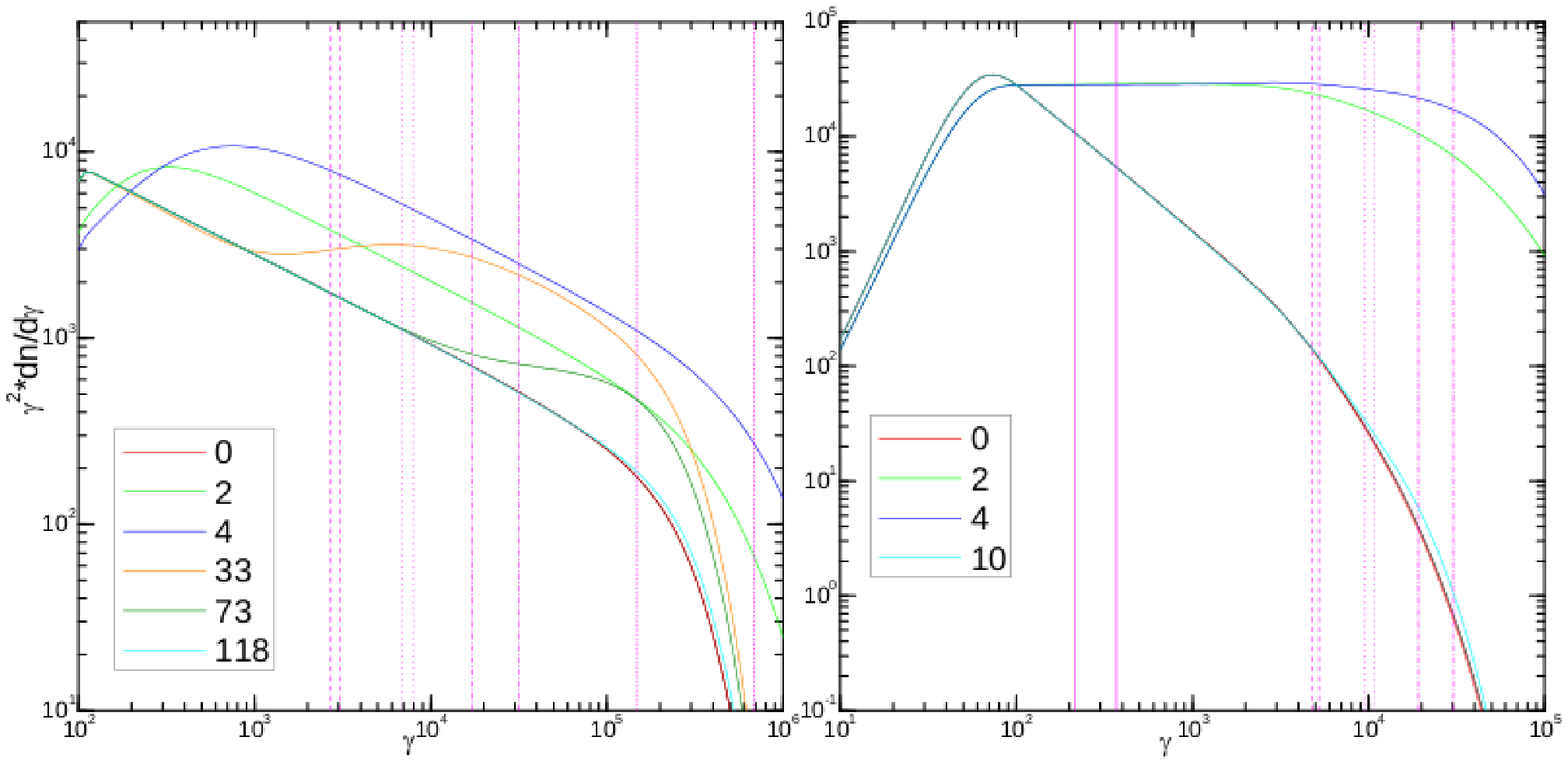}
\vspace{5pt}
\caption{Time evolution of the electron spectra for scenario 3. This case has two additional lines
(orange and olive) in Mkn~421 in the evolving period to show the hardening at lower energies and the softening
at higher energies. Otherwise panels ({\it left:} Mkn~421; {\it right:} PKS~1510-089)
and line styles are as in Fig. \ref{Electron5}.
\label{Electron4}}
\end{figure}

For Mkn~421, due to the unchanged magnetic field, the higher-energy electrons take longer
to cool than in the previous scenarios so that the flare duration is longer.
Also, the electron spectral index remains nearly constant while the shock
is present (Fig. \ref{Electron4}), so that $\Pi_{max}$ will be nearly unchanged throughout
the emission region. However, after the shock leaves a given zone, the spectrum hardens at
lower energies while softening at higher energies. Since this effect acts extremely slowly
and is very weak, its contribution to both luminosity and the polarization can
only be seen during the post-peak phase. As a result, the polarization region
dominates only because of its high luminosity. Another reason for the different polarization
behavior with respect to the previous scenarios is that the polarization region is larger,
containing more evolving zones. This effect is especially strong after the flare peak, when
much of the emission region has been affected by the shock. Consequently, the emission region
is nearly equivalent to the initial state, except that all zones radiate with higher luminosity.
Therefore, we see that the polarization percentage is lower overall than in the previous scenarios,
and the pre-peak polarization percentage is higher than in the post-peak phase. Just like the light
curves, also the polarization percentage takes longer to evolve at higher energies.

The situation for PKS~1510-089 is somewhat different. When the shock reaches a given zone, the
shortened acceleration timescale results in a much harder spectrum, which will give lower $\Pi_{max}$.
After the shock leaves the zone, due to the strong EC cooling, the electron spectrum quickly evolves back
to equilibrium (Fig. \ref{Electron4}). As a result, the polarization region is very narrow.
However, a much more significant factor is that the flare-to-equilibrium luminosity ratio is very large
in this case (Fig. \ref{Pks15104}).
Therefore, the contribution from background regions to the polarization is negligible.
Hence, although $\Pi_{max}$ is lower in the polarization region, this effect is compensated by
the highly ordered magnetic field in the polarization region in the pre-peak ($\sim\!-45^{\circ}$)
and post-peak ($\sim\!-135^{\circ}$) periods of the flare. The basin at the flare peak is, again,
due to the axisymmetry of the polarization region. The PA shows similar features, achieving its minimum
quickly at the beginning of the flare, gradually evolving to $90^{\circ}$ at the peak,
then to maximum at the end of the flare and back to $90^{\circ}$ in equilibrium. There is one exception, however:
At radio frequencies, the flare-to-equilibrium ratio is nearly $1$, thus we see both the polarization
percentage and angle staying nearly constant.

\begin{figure}[ht]
\centering
\includegraphics[width=15cm]{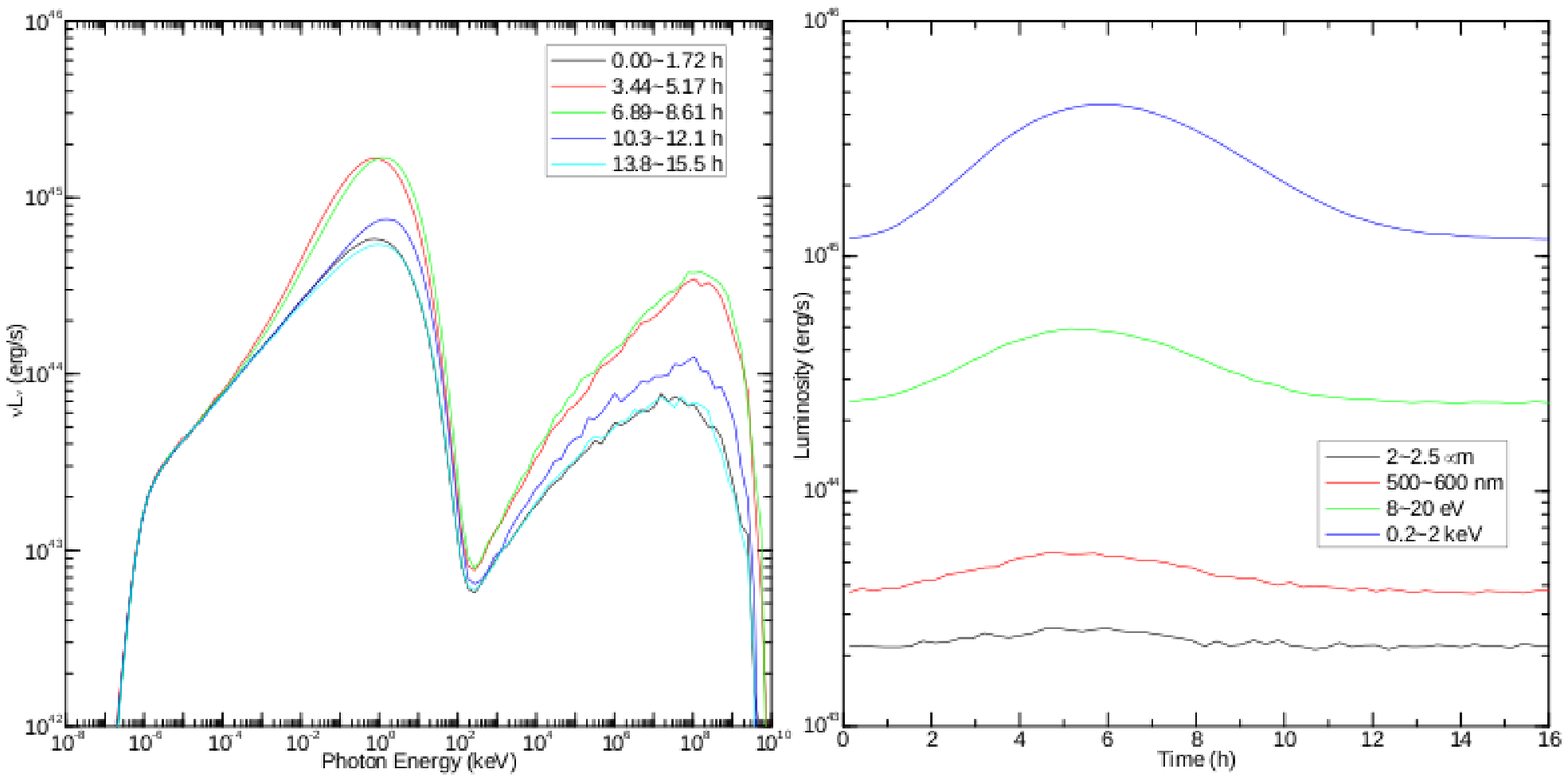}
\vspace{10pt}
\includegraphics[width=15cm]{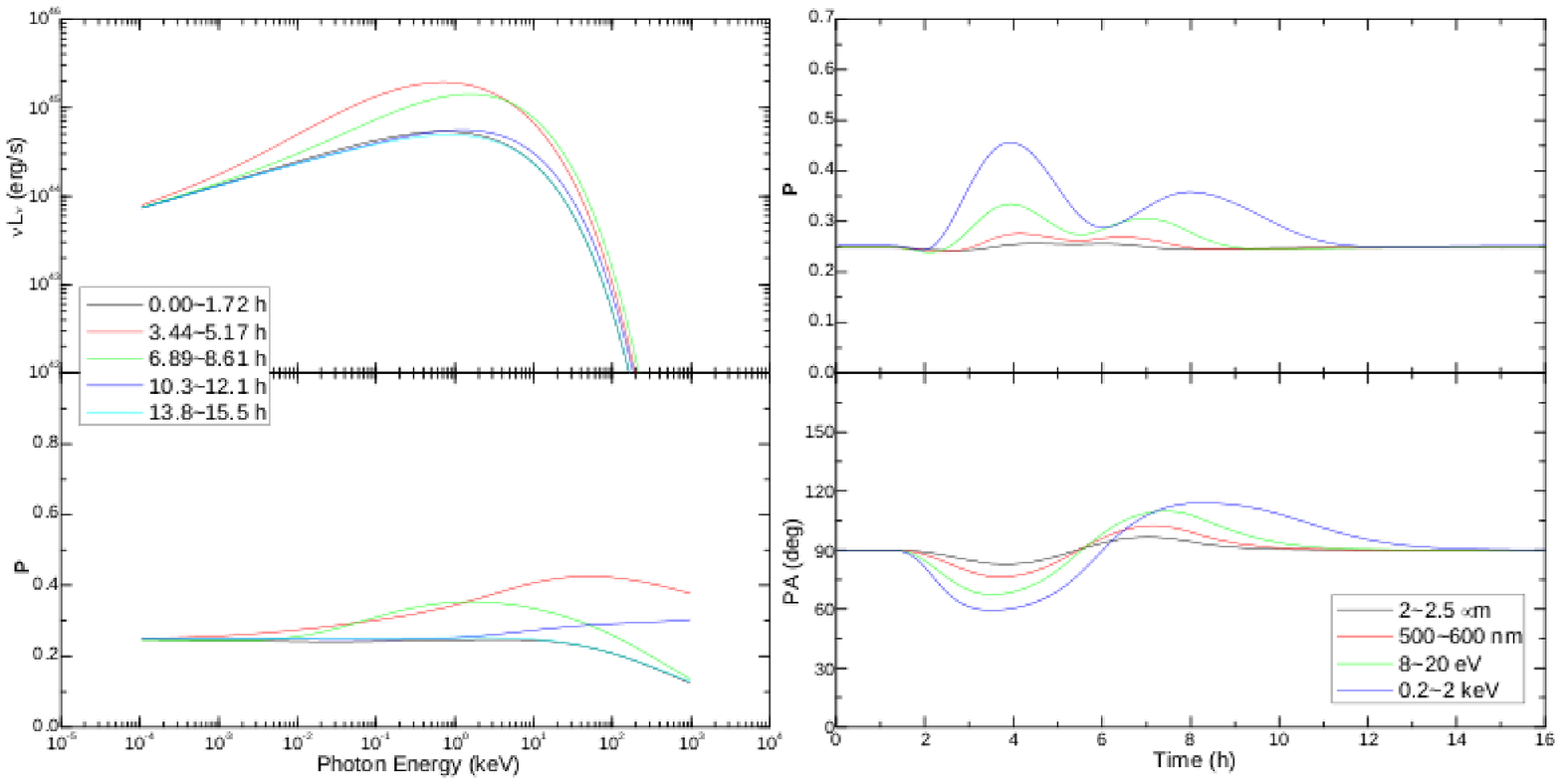}
\vspace{5pt}
\caption{Scenario 4 (injection of additional high-energy electrons) for Mkn~421. Panels and line
styles are as in Fig. \ref{Mkn4215}.
\label{Mkn4211}}
\end{figure}

\subsection{Injection of Particles}

In this scenario, the shock is assumed to continuously inject relativistic particles
in the zones that it crosses (parameters for the injected electrons
can be found in Table 1). The newly injected electrons will evolve and radiate
immediately after the injection, in the same way as the original electrons in that
zone. This scenario is similar to the previous one, except for the following differences.

\begin{figure}[ht]
\centering
\includegraphics[width=15cm]{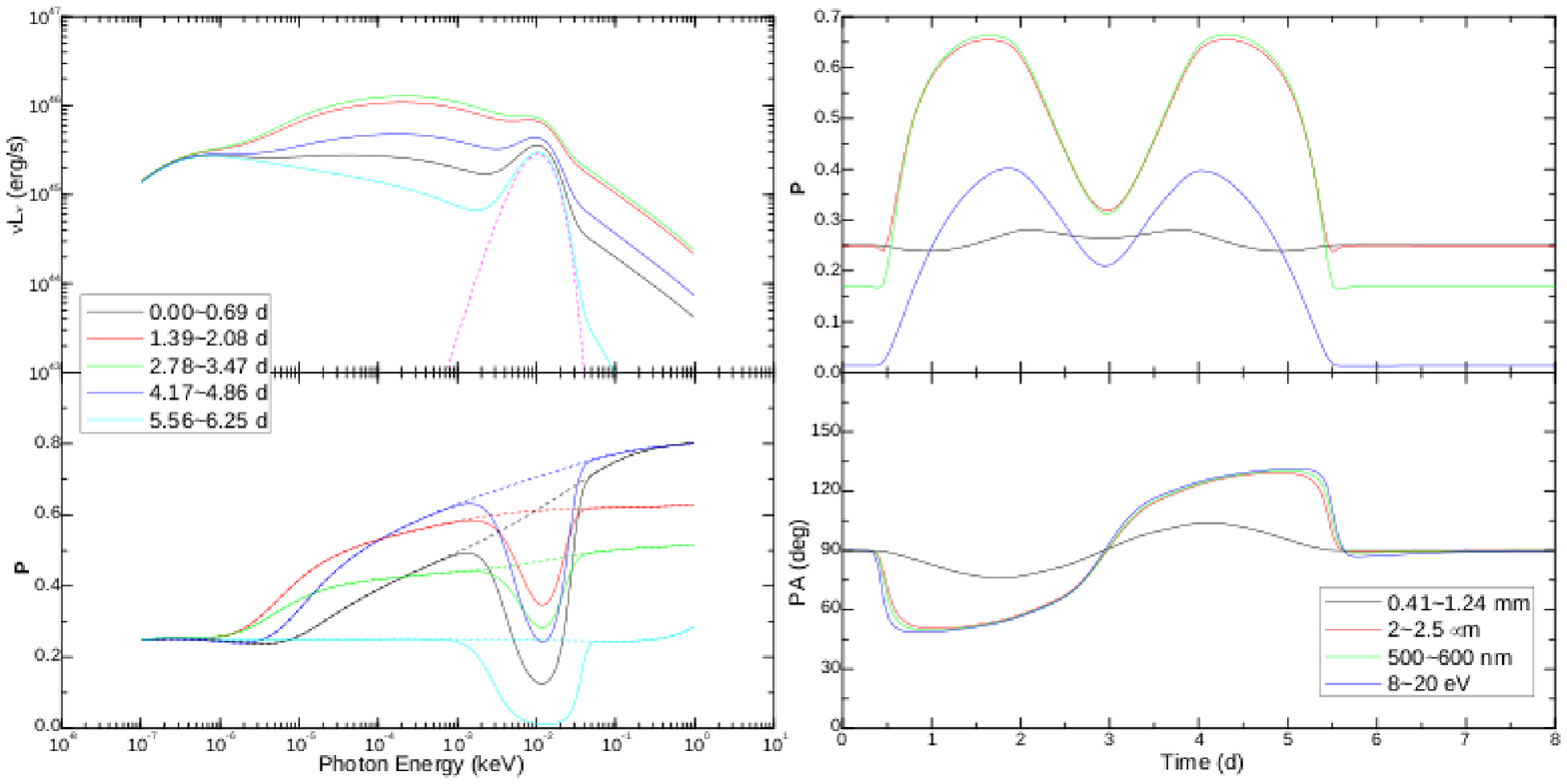}
\vspace{5pt}
\caption{Scenario 4 (injection of additional high-energy electrons) for PKS~1510-089. Panels and
line styles are as in Fig. \ref{Pks15103}.
\label{Pks15101}}
\end{figure}

\begin{figure}[ht]
\centering
\includegraphics[width=15cm]{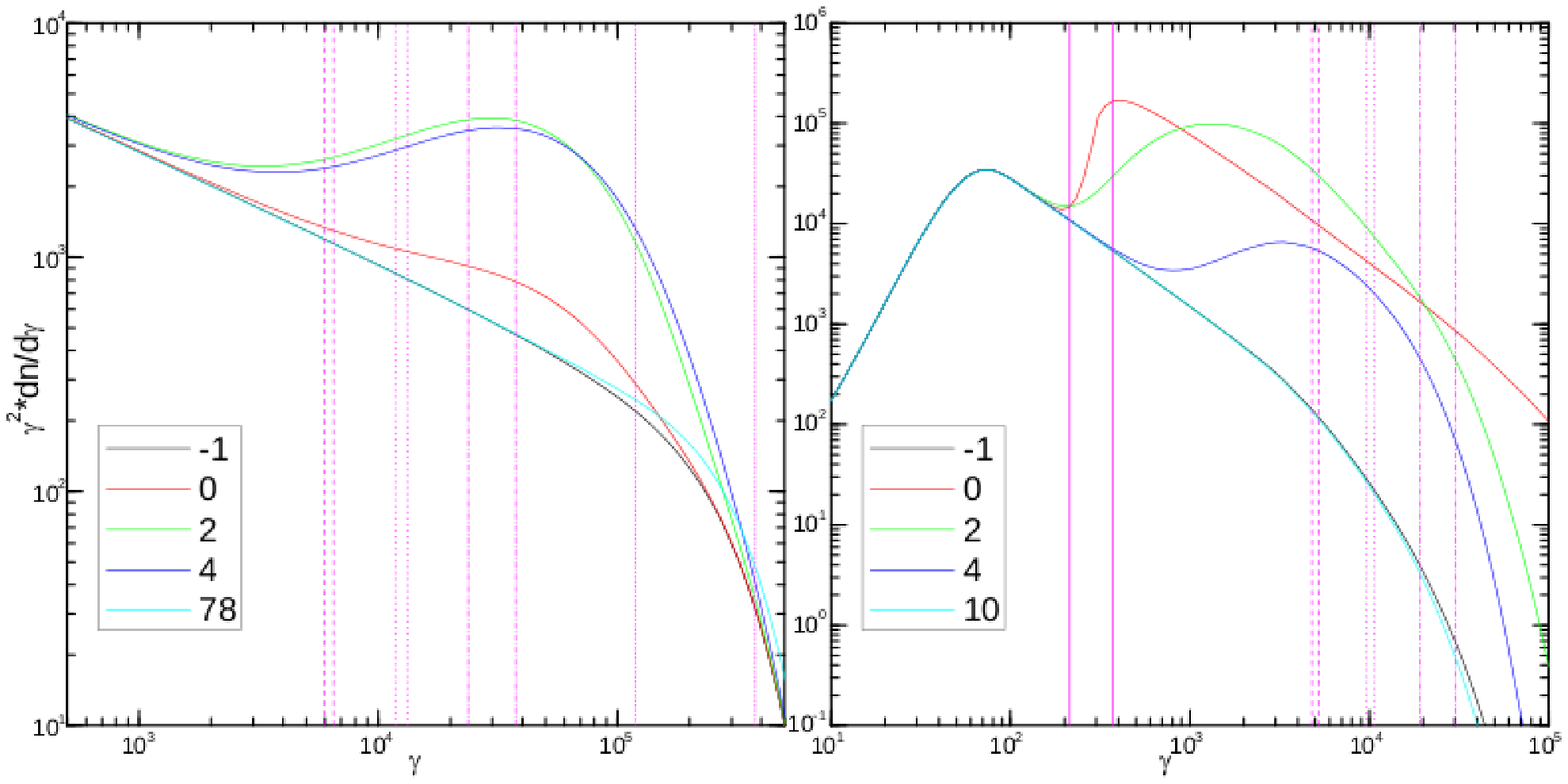}
\vspace{5pt}
\caption{Time evolution of the electron spectra for scenario 4. Notice in this case the spectra at
pre-shock equilibrium and at the onset of the shock are different, as the electrons are injected.
Panels (left: Mkn~421; right: PKS~1510-089) and line styles are as in Fig. \ref{Electron5}.
\label{Electron1}}
\end{figure}

First, in the case of Mkn~421, the newly injected electrons occupy an energy range not
extending beyond the equilibrium electron distribution (see Fig. \ref{Electron1}).
In particular, the flare electron spectrum will not extend to higher energies than the
equilibrium distribution, and therefore the electron cooling timescales remain almost unaffected.
Consequently, the X-ray flare in Mkn~421 again stops earlier (Fig. \ref{Mkn4211}).
Second, although immediately after the injection the electron spectrum hardens, at the highest
energies, it will become softer than the initial spectrum while the additional high-energy electrons
cool off to lower energies, resulting in a higher $\Pi_{max}$. However, at lower energies the flare
electron spectrum will generally be harder than the equilibrium spectrum. Therefore,
the polarization percentage in general increases at higher energies, but decreases at lower energies
(Figs. \ref{Mkn4211}, \ref{Pks15101}). The same applies to PKS~1510-089, but we observe
that the polarization percentage increases a little bit in radio but decreases in ultraviolet.
The reason is that the radio has a little bit higher flare-to-equilibrium luminosity ratio than
that in the previous scenario, while that for ultraviolet is lower, so that its polarization is
contaminated more by the external photon field in the dusty torus. The PA swings follow similar patterns as in the previous
scenario.

\section{Discussion\label{section4}}

In \S3, we have shown both the energy and the time dependencies of the synchrotron fluxes and polarization
patterns in a generic shock-in-jet scenario for 4 different possible mechanisms through which a shock may
result in synchrotron flaring behavior. We have chosen model parameter values that have been shown to be
appropriate to reproduce SEDs and light curves or Mkn~421 and PKS~1510-089. However, there are still
parameter degeneracies, and some of the geometric parameters, such as the ratio between $z$ and $r$,
and the viewing angle, i.e., the direction of the LOS with respect to the jet axis, $\theta_{\rm obs}$,
have been fixed without strict observational constraints.
In this section, we will show that the choice of these parameters may have a non-negligible influence on
the predicted polarization patterns. Specifically, we use $\theta_{\rm obs}$ as an example to discuss the
geometric effect on the polarization.

\subsection{Dependence on the viewing angle\label{section41}}

Throughout \S\ref{section3}, we have assumed that we are observing the blazar jet from the side
($\theta_{\rm obs} = 90^o$) in the co-moving frame, due to relativistic aberration. However, the
relativistic beaming effects will be very similar for viewing angles that are a few degrees off
this angle --- in particular towards smaller viewing angles. Here we investigate the scenario 1
(for which we have shown that large PA rotations are naturally predicted) under two different viewing
angles, $\theta_{\rm obs}$, namely $60^{\circ}$ and $80^{\circ}$, to illustrate this geometric effect
on the polarization. $\theta_{obs}$ is unlikely to be substantially greater than $90^{\circ}$, as
we are observing the blazar jet from within the relativistic beaming cone, given by $\theta^{\ast}_{\rm obs}
\lesssim\!1/\Gamma$ in the observer's frame, which corresponds to $\theta_{\rm obs} \lesssim 90^{\circ}$
in the comoving frame.

\begin{figure}[ht]
\centering
\includegraphics[width=15cm]{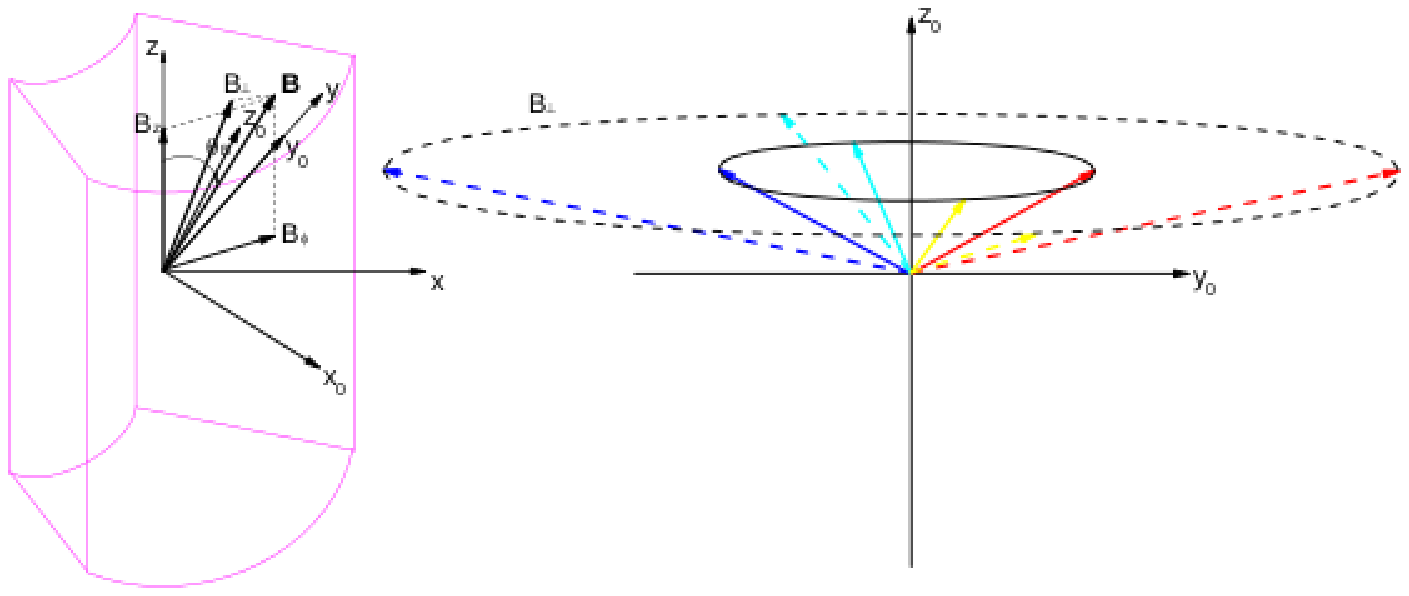}
\vspace{5pt}
\caption{Similar to Fig. \ref{bsketch1} but for $\theta_{obs}=60^{\circ}$.
Here the {\it yellow} ($+y$) and {\it cyan} ($-y$) regions are not symmetric, hence the strengths of $B_{x_0}$ and $B_{y_0}$
change accordingly.
\label{bsketch2}}
\end{figure}

With an off-side $\theta_{\rm obs}$, the polarization region will be a bit smaller and located
differently. $B_{y_0}$ is as well not affected, but the major change here is $B_{z_0}$.
We observe that $B_{\perp}$, as well as the effective poloidal component $B_{z_0}$,
is stronger in $-y$ while weaker in $+y$ (Fig. \ref{bsketch2}). Thus the axisymmetry
discussed in \S\ref{section3} is invalid. As a result, the emission from $-y$, where $B_{\perp}$
has a relatively stronger poloidal contribution, will dominate over the emission from $+y$, where
$B_{\perp}$ has a dominant toroidal contribution. However, in the initial state, although
$B_{z_0}$ is stronger near the $-y$ axis, it is much weaker near the $+y$ axis and
near the $x=\pm r_{max}$ boundaries, thus the polarization due to $B_{z_0}$ is overall
weaker than that in the $\theta_{\rm obs} = 90^{\circ}$ case. Therefore, at the pre-flare
and post-flare equilibrium states, the PA has the same value as before, while the polarization
percentage is lower.

\begin{figure}[ht]
\centering
\includegraphics[width=15cm]{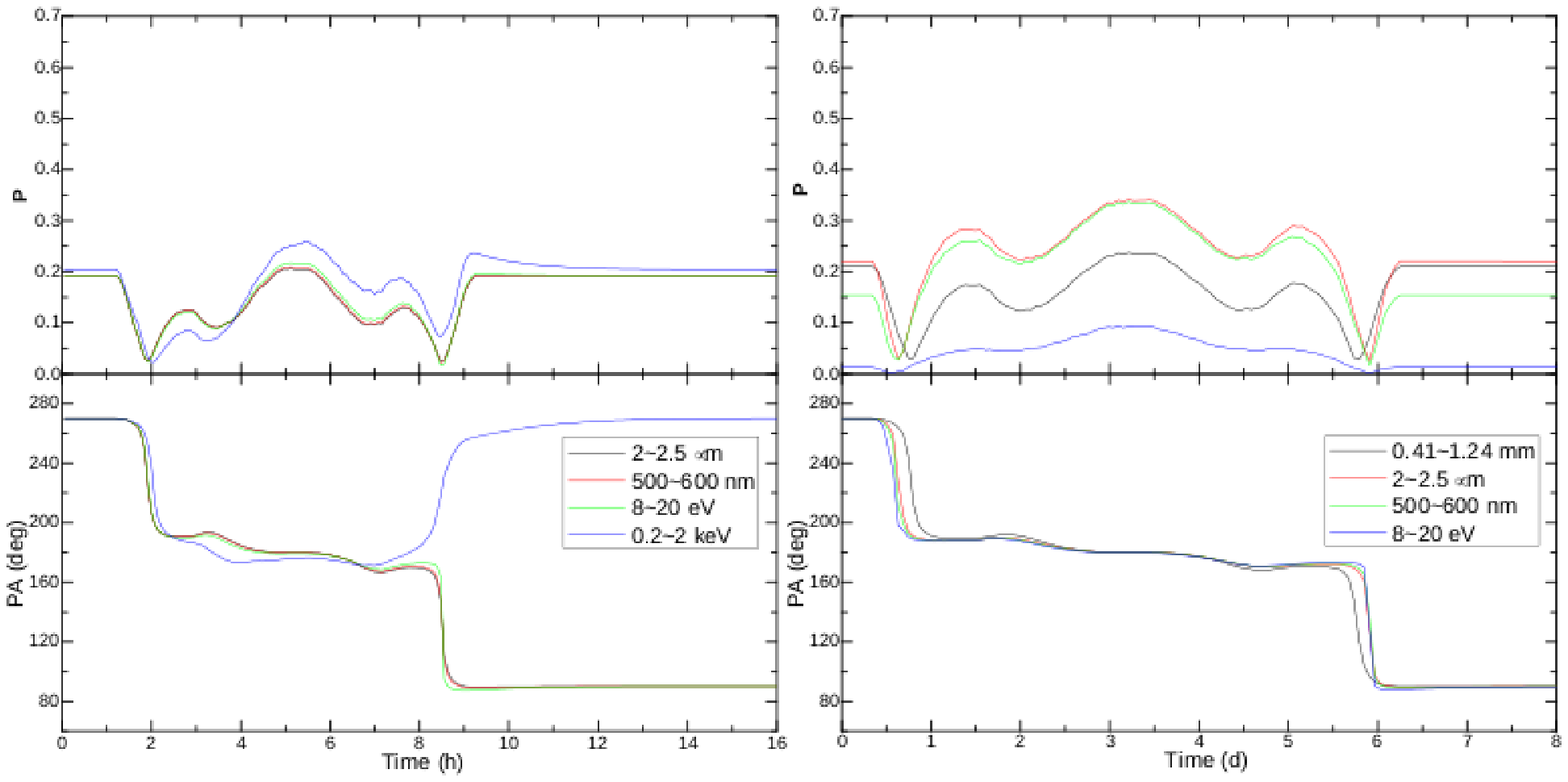}
\vspace{10pt}
\includegraphics[width=15cm]{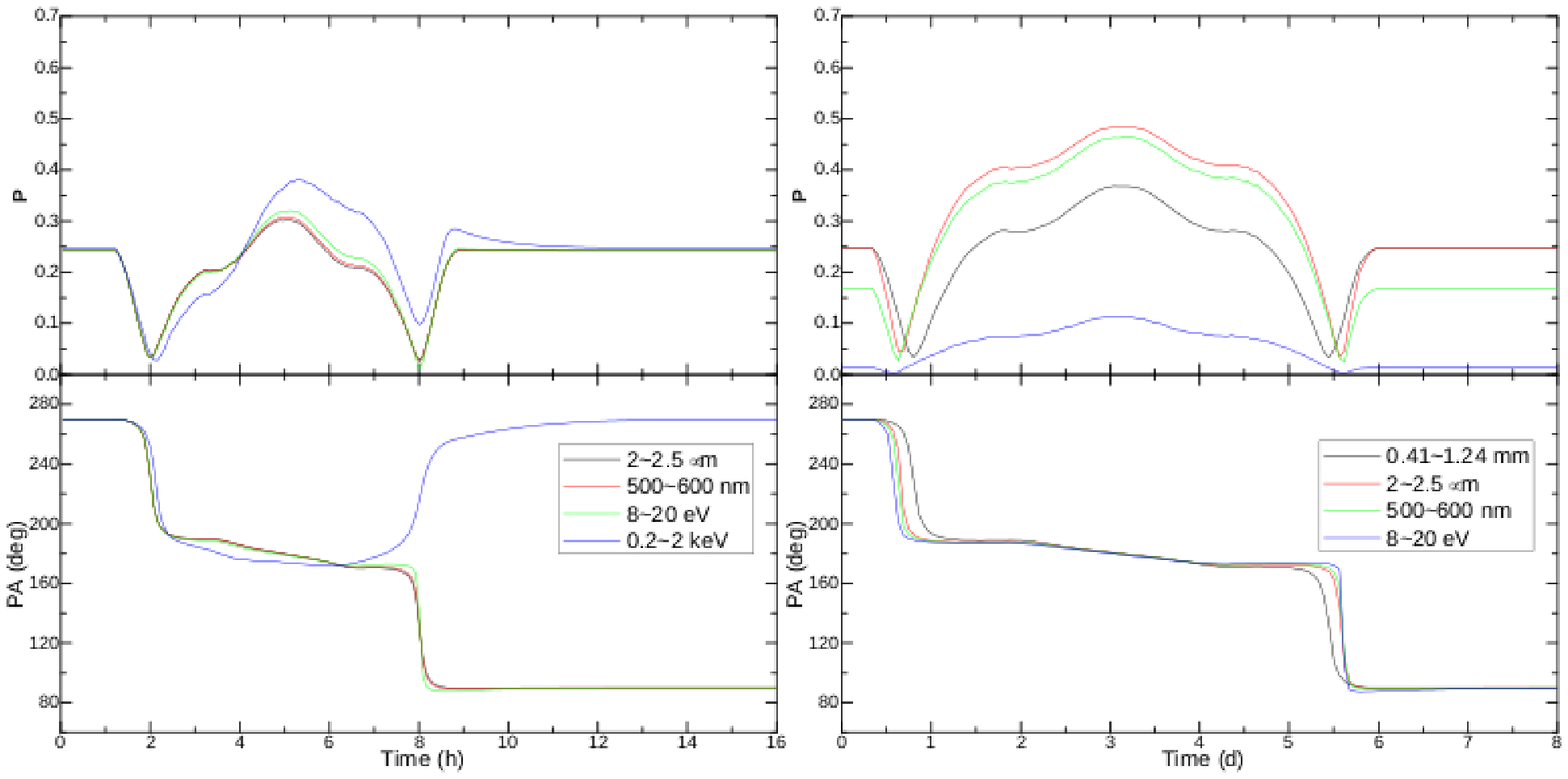}
\vspace{5pt}
\caption{Polarization vs time plots for flaring scenario 1, for different viewing angles, for
comparison with Figs. \ref{Mkn4215} and \ref{Pks15105}. {\it Left column:} Mkn~421.
{\it Right column:} PKS~1510-089. {\it Top row:} $\theta_{obs}=60^{\circ}$.
{\it Bottom row:} $\theta_{obs}=80^{\circ}$.\label{angles}}
\end{figure}

During the flare, however, unlike in \S\ref{section31} where amplification of $B_{\phi}$ leads to a dramatic
increase in $B_{y_0}$, this time it also contributes to $B_{z_0}$, especially at the flare peak (Fig. \ref{bsketch2}).
As a result, the polarization due to $B_{y_0}$ will be balanced out more by that from $B_{z_0}$. Hence the polarization
caused by the toroidal magnetic-field component takes longer to reach maximum after its dominance over the original
polarization due to $B_{z_0}$, so that the dip in the polarization percentage vs time is wider (Fig. \ref{angles}),
and the polarization percentage is generally lower with smaller $\theta_{obs}$ (obviously, the net polarization goes
to zero in the limit $\theta_{\rm obs} \to 0^{\circ}$). This effect is particularly strong in the $\theta_{obs}=60^{\circ}$
case shown in Fig. \ref{angles}, where we observe that in the pre-peak and the post-peak flaring state, there
are two small dips in the polarization percentage with corresponding fluctuations in the PA. This is because
the toroidal component is less dominant: At the beginning of the flare, the polarization region is
small, but the toroidal component is highly ordered and is oriented in the $y_0$ direction (Fig. \ref{bsketch2}).
This will give strong polarization in $PA=180^{\circ}$, which will quickly cancel out the background $PA=270^{\circ}$
polarization and dominate. However, when the polarization region moves closer to the center, $B_{z_0}$ will increase
on the $-y$ side, which dominates the emission; meanwhile, the background region will be dominated by emission from
the central and $-x$ regions, which will have stronger poloidal polarization than produced in the $+x$ region in the
initial state. Hence, the poloidal contribution to the polarization increases.
When the polarization region moves to the flare-peak position, however, the central region is affected
by the shock. Although $B_{z_0}$ will become even stronger near the $-y$ axis, $B_{\perp}$ in its neighborhood
has a stronger $B_{y_0}$ component. Since the polarization region extends to neighboring regions, the polarization
due to $B_{y_0}$ will regain its dominance. The post-peak and the post-flare equilibrium evolve in the same way,
as the polarization region is symmetric in the time domain, except for slight differences in X-ray.

\section{Summary and Conclusions}

In this paper, we have presented a detailed analysis of time- and energy-dependent synchrotron
polarization signatures in a shock-in-jet model for $\gamma$-ray blazars. Our calculations employ
a full 3D radiation transfer code, assuming a helical magnetic field throughout the jet, carefully
taking into account light-travel-time and all other relevant geometric effects. We considered
several possible mechanisms through which a relativistic shock propagating through the jet may affect
the jet plasma to produce a synchrotron and high-energy flare. Among the scenarios investigated, we
found that a compression of the magnetic field, increasing the toroidal field component and thereby
changing the direction of the magnetic field in the region affected by the shock, leads to correlated
synchrotron + SSC flaring, associated with substantial variability in the synchrotron polarization
percentage and position angle. Most importantly, this scenario naturally explains large PA rotations
by $\gtrsim 180^o$, as observed in connection with $\gamma$-ray flares in several blazars. In particular,
we have falsified the claim \citep[e.g.,][]{Abdo10} that pattern propagation through an axisymmetric,
straight jet can not produce large PA swings and rotations.

Alternative models to explain polarization variability and PA rotations, include a helical guiding
magnetic field, which forces plasmoids to move along helical paths \citep{VR99}, and the TEMZ model
by \cite{Marscher14}. \cite{Abdo10} have suggested that PA swings correlated with $\gamma$-ray flaring
activity may result when a shock or other disturbance propagates along a curved (helical) jet. In the
course of the propagation along a curved trajectory, the observer's viewing angle with respect to
the co-moving magnetic field in the active region changes, leading to possible PA swings. While
such an explanation seems plausible on geometric grounds, no quantitative analysis of the resulting,
correlated synchrotron and high-energy flux and polarization features has been presented for such
a model, and there is currently no evidence (e.g., from observations or from MHD simulations)
that blazar jets are guided by sufficiently strong, helical magnetic fields that would be able
to guide relativistic pattern propagation along helical trajectories. Our analysis in this paper
has demonstrated that light-travel time effects lead to much more complicated time-dependent
polarization features than predicted by purely geometric considerations that neglect LTTEs.

By the stochastic nature of the TEMZ model \citep{Marscher14}, it predicts generally asymmetrical
light curves and random polarization patterns that do only occasionally (by coincidence) lead to
large-angle PA swings, which will generally not be correlated with pronounced flaring activity at
higher energies. Observed polarization angle changes do, in fact, often appear stochastic in nature,
and even the polarization-swing event reported in \cite{Abdo10} showed signs of non-unidirectional
PA changes and may therefore be interpreted by a stochastic model such as the TEMZ model.
The TEMZ code of \cite{Marscher14} takes into account SSC scattering (and its influence on
electron cooling) only with seed photons from the central mach disk. Therefore, it is well
applicable for blazars in which $\gamma$-ray emission and electron cooling are dominated by
Comptonization of external radiation fields, which appears to be the case in low-frequency
peaked blazars (FSRQs, LBLs), but not for HBLs like Mrk~421, in which the $\gamma$-ray emission
is well modeled as being dominated by SSC radiation.

The strength of the PA rotation model presented here is that it very naturally explains large
PA rotations, correlated with $\gamma$-ray flaring events, without the need for non-axisymmetric
jet features. It is supported by observations of large-angle, uni-directional polarization swings,
e.g., in 3C279 \citep{Kiehlmann13}, which suggest that such features are unlikely to be caused by
a stochastic process, but are likely the result of preferentially ordered structures. For these
reasons, we prefer our quite natural explanation of PA swings correlated with synchrotron and
high-energy flares, resulting from light-travel-time effects in a shock-in-jet model in a straight,
axisymmetric jet embedded in a helical magnetic field.

\acknowledgments{We thank Alan Marscher for valuable discussions and comments on this manuscript.
This work was supported by NASA through Fermi Guest Investigator Grant no. NNX12AP20G.
HZ is supported by the LANL/LDRD program and by DoE/Office of Fusion Energy
Science through CMSO. XC acknowledges support by the Helmholtz
Alliance for Astroparticle Physics HAP funded by the Initiative and Networking
Fund of the Helmholtz Association. XC gratefully acknowledges the support
during his visit to LANL when this work was started.
MB acknowledges support by the South African Research Chairs Initiative of the Department of Science and
Technology and the National Research Foundation of South Africa.
Simulations were conducted on LANL's Institutional Computing machines.}

\clearpage

\end{document}